\documentclass[fleqn,usenatbib]{mnras}

\usepackage{newtxtext,newtxmath}
\usepackage[T1]{fontenc}

\DeclareRobustCommand{\VAN}[3]{#2}
\let\VANthebibliography\thebibliography
\def\thebibliography{\DeclareRobustCommand{\VAN}[3]{##3}\VANthebibliography}

\usepackage{graphicx}	% Including figure files
\usepackage{amsmath}	% Advanced maths commands
\usepackage{listings}
\usepackage{siunitx}
\usepackage{subcaption}
\usepackage{hyperref}
\usepackage{lipsum}
\newcommand{\angstrom}{\text{\normalfont\AA}}

\newcommand{\ba}{\[\begin{aligned}}
\newcommand{\ea}{\end{aligned}\]}
\usepackage{colortbl}
\newcommand{\eq}[1]{\begin{align}#1\end{align}}

\makeatletter
\g@addto@macro\bfseries{\boldmath}
\makeatother

\DeclareSIUnit \pkpc {pkpc}
\DeclareSIUnit \parsec {pc}
\DeclareSIUnit \h {h}
\DeclareSIUnit\year{yr}

\newcommand{\lya}{Ly$\alpha$}
\newcommand{\fesc}{$f_{\rm esc}^{\rm LyC}$}
\newcommand{\fesclos}{$f_{\rm esc,\;LOS}^{\rm LyC}$}
\newcommand{\fesca}{$f_{\rm esc}^{{\rm Ly}\alpha}$}
\newcommand{\wlya}{$W_{\lambda}({\rm Ly}\alpha)$}
\newcommand{\vsep}{$v_{\rm sep}$}

\newcommand{\fcen}{$f_{\rm cen}$}
\newcommand{\asym}{$A_f^{\rm red}$}
\newcommand{\sphinx}{{\small SPHINX$^{20}$}}
\newcommand{\fx}{$f_{\rm esc}^{\rm LyC} \xi_{\rm ion}$}

\newcommand*{\lb}  {\left(}
\newcommand*{\rb}  {\right)}
\newcommand*{\ls}  {\left[}
\newcommand*{\rs}  {\right]}

\title[\lya\ and LyC Escape from JWST Analogues]{The Great Escape: Understanding the Connection Between \lya\ Emission and LyC Escape in Simulated JWST Analogues}

\author[N. Choustikov, et al.]{Nicholas Choustikov$^{1}$\thanks{nicholas.choustikov@physics.ox.ac.uk},
Harley Katz$^{2,1}$, 
Aayush Saxena$^{1, 3}$,
Thibault Garel$^{4}$,
Julien Devriendt$^{1}$,\newauthor
Adrianne Slyz$^{1}$,
Taysun Kimm$^{5}$,
Jeremy Blaizot$^{6}$, and
Joki Rosdahl$^{6}$
\\
$^{1}$Sub-department of Astrophysics, University of Oxford, Keble Road, Oxford OX1 3RH, United Kingdom\\
$^{2}$Department of Astronomy \& Astrophysics, University of Chicago, 5640 S Ellis Avenue, Chicago, IL 60637, USA\\
$^{3}$Department of Physics and Astronomy, University College London, Gower Street, London WC1E 6BT, United Kingdom\\
$^{4}$Observatoire de Gen\`{e}ve, Universit\'{e} de Gen\`{e}ve, Chemin Pegasi 51, 1290 Versoix, Switzerland\\
$^{5}$Department of Astronomy, Yonsei University, 50 Yonsei-ro, Seodaemun-gu, Seoul 03722, Republic of Korea\\
$^{6}$CNRS, Centre de Recherche Astrophysique de Lyon UMR5574, Univ Lyon, Univ Lyon1, Ens de Lyon, F-69230 Saint-Genis-Laval, France\\
}

\date{Accepted XXX. Received YYY; in original form ZZZ}

\pubyear{2024}

\begin{document}
\label{firstpage}
\pagerange{\pageref{firstpage}--\pageref{lastpage}}
\maketitle

% Abstract of the paper
\begin{abstract}
Constraining the escape fraction of Lyman Continuum (LyC) photons from high-redshift galaxies is crucial to understanding reionization. Recent observations have demonstrated that various characteristics of the Ly$\alpha$ emission line correlate with the inferred LyC escape fraction ($f_{\rm esc}^{\rm LyC}$) of low-redshift galaxies. Using a data-set of 9,600 mock Ly$\alpha$ spectra of star-forming galaxies at $4.64 \leq z \leq 6$ from the {\small SPHINX$^{20}$} cosmological radiation hydrodynamical simulation, we study the physics controlling the escape of Ly$\alpha$ and LyC photons. We find that our mock Ly$\alpha$ observations are representative of high-redshift observations and that typical observational methods tend to over-predict the Ly$\alpha$ escape fraction ($f_{\rm esc}^{\rm Ly\alpha}$) by as much as two dex. We investigate the correlations between $f_{\rm esc}^{\rm LyC}$ and $f_{\rm esc}^{\rm Ly\alpha}$, Ly$\alpha$ equivalent width ($W_{\lambda}({\rm Ly\alpha})$), peak separation ($v_{\rm sep}$), central escape fraction ($f_{\rm cen}$), and red peak asymmetry ($A_f^{\rm red}$). We find that $f_{\rm esc}^{\rm Ly\alpha}$ and $f_{\rm cen}$ are good diagnostics for LyC leakage, selecting for galaxies with lower neutral gas densities and less UV attenuation that have recently experienced supernova feedback. In contrast, $W_{\lambda}({\rm Ly\alpha})$ and $v_{\rm sep}$ are found to be necessary but insufficient diagnostics, while $A_f^{\rm red}$ carries little information. Finally, we use stacks of Ly$\alpha$, H$\alpha$, and F150W mock surface brightness profiles to find that galaxies with high $f_{\rm esc}^{\rm LyC}$ tend to have less extended Ly$\alpha$ and F150W haloes but larger H$\alpha$ haloes than their non-leaking counterparts. This confirms that Ly$\alpha$ spectral profiles and surface brightness morphology can be used to better understand the escape of LyC photons from galaxies during the Epoch of Reionization.
\end{abstract}


%%%%%%%%%%%%%%%%%%%%%%%%%%%%%%%%%%%%%%%%%%%%%%%%%%

%%%%%%%%%%%%%%%%% BODY OF PAPER %%%%%%%%%%%%%%%%%%

\section{Introduction}
\label{sec:introduction}

The Universe transitioned from a predominantly neutral to ionized state by the redshift interval of $5\lesssim z\lesssim6$ \citep{Planck:2018,Kulkarni:2019a,Keating:2020, Becker:2021, Bosman:2022}. However, the beginning of this process of reionization, the nature of the sources responsible, and the evolution of the neutral fraction all still remain uncertain. Observational upper and lower limits have been placed on the first of these questions by inferring star formation histories (SFHs) of high-redshift galaxies \citep[e.g.][]{Laporte:2021} as well as by the Cosmic Microwave Background \citep[e.g.][]{Heinrich:2021}. In contrast, model-dependent constraints have been derived for the evolution of the neutral fraction from the damping wings of high-redshift quasars \citep{Davies:2018,Greig:2019,Durovcikova:2020}, the decrease in number densities of Lyman-$\alpha$ (\lya) emitters (LAEs) \citep{Stark:2010, Pentericci:2011, Mason:2018, Jones:2023} as well as the opacity of the \lya\ forest \citep{Fan:2006b, Fan:2006}.

There remains debate about the sources of ionising photons needed to drive this period of reionization. Generally, due to number density arguments derived from the steep, faint-end slope of the high-redshift UV luminosity function \citep[e.g.][]{Bouwens:2015}, the primary candidates tend to be low-mass galaxies. Of these, the star-forming galaxies (SFGs) are typically considered due to the amounts of Lyman Continuum (LyC) photons they produce \citep[e.g.][]{Robertson:2015}. Beyond this however, other properties (e.g. mass, metallicity, sizes, morphology) of such galaxies remain largely unknown. This is important as the source model impacts the topology of reionization. Specifically, modifying the shape and amplitude of the 21-cm signal \citep{Zaldarriaga:2004, McQuinn:2007, Kulkarni:2017}, as well as impacting the subsequent evolution of these galaxies themselves \citep[e.g.][]{Efstathiou:1992,Shapiro:2004,Gnedin:2014,Rosdahl:18,Katz:2020,Ocvirk:2020,Bird:2022,Borrow:2023}. Beyond this, some of the ionising photon budget is likely to be contributed by active galactic nuclei \citep[AGN, e.g.][]{Madau:2015, Chardin:2017, Torres-Alba:2020} or even more extreme, rare sources, including mini-quasars \citep{Haiman:1999, Ricotti:2004, Hao:2015} and shocks \citep{Dopita:2011, Wyithe:2011}. While all of these sources can produce a large number of ionising photons, their rarity and possible dust obscuration limits their relative contribution. In the case of AGN, they are only considered to dominate the ionizing budget at $z\lesssim 4$ \citep[e.g.][]{Kulkarni:2019b,Dayal:2020,Trebitsch:2021}.

In all cases, the production rate of LyC photons for a given source is given by the UV luminosity function ($\rho_{\rm UV}$), LyC photon production rate per UV luminosity ($\xi_{\rm ion}$) and the LyC photon escape fraction ($f^{\rm LyC}_{\rm esc}$), defined as the fraction of LyC photons that escape the virial radius of their host galaxy. Both $\rho_{\rm UV}$ and $\xi_{\rm ion}$ have been fairly well constrained, due to the fact that $\rho_{\rm UV}$ can be measured from deep imaging surveys \citep[e.g.][]{Bowler:2020,Harikane:2022,Donnan:2023,Varadaraj:2023}, while $\xi_{\rm ion}$ can be predicted by stellar population synthesis models \citep[e.g.][]{Leitherer:1999, Stanway:2018} or inferred from Balmer line emission \citep[e.g.][]{Maseda:2020,Saxena:2023}. Uncertainties in $\xi_{\rm ion}$ are driven by differences in these models themselves (e.g. binaries, IMF, gas geometry etc.). In contrast, the escape fraction is much more difficult to measure directly. This is due to the fact that it is sensitive to complex non-linear physics in the interstellar medium (ISM), can be highly line-of-sight dependent, and can not be directly measured at redshifts $\gtrapprox 4$ due to the increasingly neutral intergalactic medium \citep[IGM, e.g.][]{Worseck:2014, Inoue:2014}. As a result, studies of the LyC escape fraction at redshifts relevant to the epoch of reionization rely on indirect measurements, made using calibrated diagnostics. These diagnostics include investigating \lya\ forest transmission through galaxies \citep[e.g.][]{Kakiichi:2018} and studies of low-redshift `analogue' galaxies \citep[e.g.][]{Leitherer:2016,Schaerer:2016,Izotov:2018}. Recently, the Low Redshift Lyman Continuum Survey \citep[LzLCS,][]{Flury:2022a, Flury:2022b} has pushed the frontier of low-redshift studies of LyC escape by observing 35 galaxies with confirmed LyC emission, nearly tripling the number of observed LyC leakers. However, the question of whether these `analogue' galaxies are representative of high-redshift galaxies is still debated \citep{Katz:2022,Katz:2023b,Schaerer:2022,Brinchmann:2023}. Furthermore, observational studies of low-redshift leakers are only able to see line-of-sight LyC emission, which is not necessarily correlated with global (i.e. angle-averaged) LyC emission \citep{Paardekooper:2015,Fletcher:2019}: the quantity which is crucial to reionization. As a result, it is useful to also model LyC escape with cosmological radiation hydrodynamic simulations \citep[e.g.][]{Gnedin:2008,Wise:2009,Kimm:2014,Xu:2016,Trebitsch:2017,Rosdahl:18, Rosdahl:22}. These simulations allow us to gain an insight into how such processes occur in high-redshift galaxies, while being informed by observational studies.

While a large number potential diagnostics have been suggested to infer LyC escape fractions (see e.g. \cite{Nakajima:2014,Jaskot:2014,Verhamme:2017,Izotov:2018,Chisholm:2018,Chisholm:2022,Flury:2022b,Saldana-Lopez:2022,Choustikov:23} and references therein), some of the most accurate metrics involve \lya\ emission. Due to the fact that \lya\ is sensitive to the neutral H~{\small I} opacity of its host galaxy \citep[e.g.][]{Verhamme:2015} and neutral H~{\small I} is the primary absorber of LyC photons, it follows that \lya\ emission can contain information about the leakage of ionizing photons. Furthermore, due to the fact that \lya\ is a resonant transition \citep{Dijkstra:2017}, the emergent spectrum holds a significant amount of information about the source \citep[e.g.][]{Orlitova:2018}, the intervening medium \citep[e.g.][]{Verhamme:2017}, as well as the geometry of the system \citep[e.g.][]{Blaizot:2023}. 

Various properties of \lya\ profiles have been suggested which trace ionising photon production as well as physical conditions in the ISM favourable to LyC leakage. These include the \lya\ equivalent width \citep[$W_{\lambda}(\rm Ly\alpha)$, e.g.][]{Henry:2015,Yang:2017,Steidel:2018, Pahl:2021}, \lya\ peak separation velocity \citep[$v_{\rm sep}$, e.g.][]{Verhamme:2015, Izotov:2018, Izotov:2021}, the asymmetry of the red peak \citep[$A_f^{\rm red}$, e.g.][]{Kakiichi:2021, Witten:2023}, the central escape fraction \citep[$f_{\rm cen}$,][]{Naidu:2022}, and the \lya\ escape fraction \citep[$f_{\rm esc}^{\rm Ly\alpha}$, e.g.][]{Dijkstra:2016, Verhamme:2017, Izotov:2020}. This diversity of potential diagnostics highlights the necessity to understand the connection between the physics of the \lya\ line and $f_{\rm esc}^{\rm LyC}$. 

Numerical simulations have been used to study this interplay, including those of isolated galaxies \citep[e.g.][]{Verhamme:2012, Behrens:2014}, \lya\ nebulae \citep[e.g.][]{Trebitsch:2016, Gronke:2017b}, molecular clouds \citep[e.g.][]{Kimm:2019, Kimm:2022, Kakiichi:2021}, zoom-in simulations of individual galaxies \citep[e.g.][]{Laursen:2019}, and large box simulations with comparatively low resolution \citep[e.g.][]{Ocvirk:2020, Gronke:2021}. \cite{Maji:2022} studied the connections between intrinsic and escaped integrated \lya\ luminosities for {\small SPHINX} galaxies and their LyC escape fractions, particularly finding a correlation between \fesca\ and $f_{\rm esc}^{\rm LyC}$. However, this work was limited by not studying the line-of-sight variability in these quantities, as well as by focusing only on these two properties of \lya\ emission. Building on this, \cite{Yuan:2024} used the {\small PANDORA} suite of high-resolution zoom-in cosmological simulations \citep{Martin-Alvarez:2023} to investigate \lya\ as a tracer of feedback-regulated LyC escape from a dwarf galaxy with varying physical models. It was concluded that there is a universal correlation between \lya\ spectral shape parameters and \fesc\ for a high time-cadence set of post-processed mock \lya\ observations from a large number of sight-lines. This reiterates our need to explore the connection between \lya\ emission and LyC escape in a statistical sample of simulated epoch of reionization galaxies, while further exploring the physics underpinning each potential diagnostic.

The aim of this present work is to investigate the line-of-sight escape and dust-attenuation of LyC and \lya\ photons from a statistical sample of \sphinx\ galaxies \citep{Rosdahl:22}, inspired by the representative sample of star-forming high-redshift galaxies with well-resolved ISMs. Using the framework presented in \cite{Choustikov:23} to test potential diagnostics for the global LyC escape fraction, we explore the physics controlling the escape of \lya\ photons and quantify the efficacy of various Ly$\alpha$-based diagnostics in predicting the LyC escape fraction.

This work is organized as follows. In Section \ref{sec:methods} we outline the numerical methods needed to simulate \lya\ emission from and transmission through \sphinx\ galaxies. Section \ref{sec:lya_props} presents and contextualises the \lya\ properties of \sphinx\ galaxies. In Section \ref{sec:results} we use our framework to contextualise and explain known \lya\ diagnostics and explore the use of extended \lya\ haloes as a potential diagnostic for the LyC escape fraction in stacked samples. Finally, in Section \ref{sec:caveats} we discuss caveats in our analysis of \lya\ radiative transfer before concluding in Section \ref{sec:conclusion}.

\section{Numerical Simulations}
\label{sec:methods}

\begin{figure*}
    \includegraphics{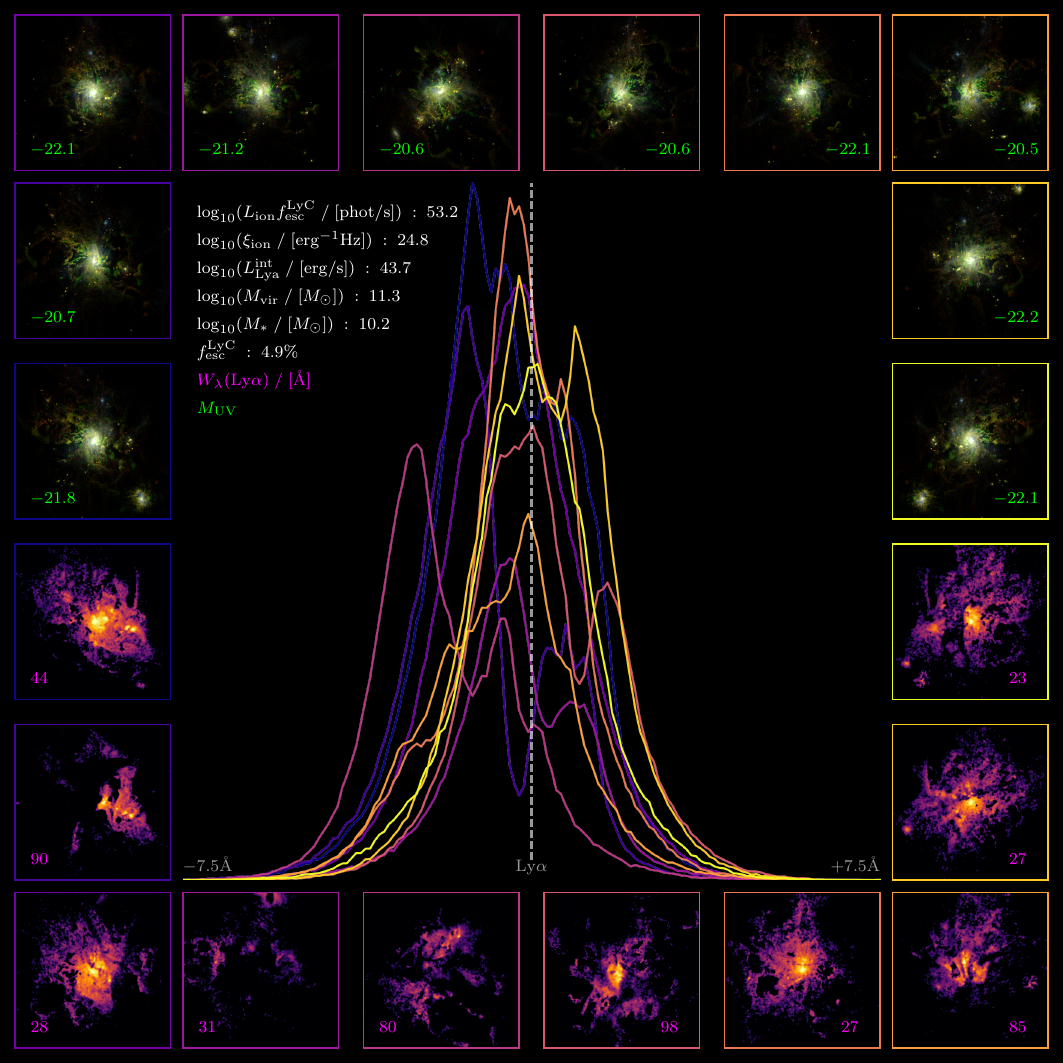}
    \caption{Complete mock data mosaic for the \protect\sphinx\ galaxy with the highest escaping LyC luminosity at $z = 5$. Dust-attenuated \protect\lya\ spectra (\textit{middle}), three colour composite images (\textit{top border}) and dust-attenuated \protect\lya\ images (\textit{bottom border}) are shown for all ten sight-lines used, indicated by the colours used. The RGB images are made using JWST filters: F365W, F444W, and F410M for the red channel, F200W and F277W for the green, while F115W and F150W are used for the blue channel (see also \protect\cite{Bezanson:2022} and \protect\cite{spdrv1}).}
    \label{fig:hero}
\end{figure*}

There is a high degree of complexity to the physical processes underlying the production and escape of Lyman $\alpha$~(\lya\ $\sim 1215.67$\angstrom) emission from galaxies. Therefore, we choose to employ state-of-the-art high-resolution numerical simulations of galaxy formation to study the production of Ly$\alpha$ photons, as well as the resonant line physics involved in their escape. We employ the \sphinx\ cosmological radiation hydrodynamics simulation \citep{Rosdahl:22}. This simulation is ideal for such a study as the volume ($20^3 \;\rm{cMpc}^3$) is large enough to capture a diversity of galaxies and it has enough resolution to model atomic cooling haloes (which contribute to LyC escape for reionization) as well as the multi-phase ISM of these systems. Full details of the simulations are provided in \cite{Rosdahl:18,Rosdahl:22}. We use mock observations from the {\small SPHINX} Public Data Release v1 \citep[SPDRv1, ][]{spdrv1}: a sample of 960 galaxies at redshifts $z \in {4.64, 5, 6}$ with $10 \;{\rm Myr-}$averaged star formation rates (SFRs) $\geq 0.3~{\rm M}_{\odot}{\rm yr}^{-1}$, representative of those observable by JWST (see discussions in Sections 3.2 and 3.3.7 of \citealt{spdrv1}, as well as \citealt{Choustikov:23}).

In order to generate the intrinsic emission of \lya\ photons for each gas cell, we use the non-equilibrium hydrogen ionization fraction directly from the simulation. For all gas cells not containing unresolved Stromgren spheres, we then calculate the exact recombination and collisional components to \lya\ emissivities for each gas cell. For gas cells with unresolved Stromgren spheres, we run a grid of spherical models using {CLOUDY v17.03} \citep{Ferland_2017}, iterated to convergence across a variety of gas densities, metallicities, ionization parameters, and electron fractions. The total intrinsic luminosity of each galaxy is then the integral over all gas cells within the virial radius. For a deeper discussion of the method used, see discussions in \cite{Choustikov:23} and \cite{spdrv1}.

Due to the fact that \lya\ is a resonant line, in order to study its propagation through an attenuating, dusty medium it is crucial to capture the diffusion of \lya\ photons both spatially and in frequency space as they escape their host galaxy. This physics is simulated using the {\small RASCAS} Monte-Carlo radiative transfer simulation code \citep{Michel-Dansac:20}. As dust evolution is not self-consistently modelled in \sphinx, we use the phenomenological Small Magellanic Cloud (SMC) dust model from \cite{Laursen:09} to assign dust to each gas cell based on their neutral gas fraction and metallicity. We also use dust asymmetry and albedo properties taken from the SMC dust model of \cite{Weingartner:2001}. The propagation of these \lya\ photons is followed until they pass the virial radius, at which point they are considered to have escaped into the intergalactic medium. Finally, as \lya\ escape is a line-of-sight sensitive process, we use the peeling algorithm \citep{Yusef-Zadeh:1984, Zheng:2002, Dijkstra:2017} to compute the \lya\ spectrum for ten different sight lines. Due to the fact that the cost of the code scales with the number of sight-lines, we select ten as a good balance between capturing the line-of-sight-driven uncertainty and computational cost. Recently, the angular dependence of these relations has been studied in more depth \citep{Blaizot:2023, Yuan:2024, Giovinazzo:2024}. In the end, this results in a total sample of 9,600 mock \lya\ spectra and images. We also utilize both global and line-of-sight LyC escape fractions, computed for photons with wavelengths $\lambda < 912\angstrom$ and $900\angstrom < \lambda < 912\angstrom$ respectively. The dust-attenuated SED for each line-of-sight is produced using similar methods (as detailed in \citealt{spdrv1} and \citealt{Choustikov:23}).

Figure \ref{fig:hero} shows the dust-attenuated \lya\ spectra (\textit{middle}), composite three-colour images (\textit{top border}) and dust-attenuated \lya\ images (\textit{bottom border}) for all ten lines-of-sight of the \sphinx\ galaxy with the highest escaping LyC luminosity at $z = 5$. The composite images are produced using JWST filters in the same way as the UNCOVER mosaic \citep{Bezanson:2022}. Namely, F365W, F444W, and F410M are used for the red channel, F200W, and F277W are used for green and F115W and F150W are used for blue \citep[see also Figure 1 of][]{spdrv1}. Border and line colours correspond to the same given line-of-sight. This demonstrates the fact that different sight-lines produce drastically different \lya\ spectra and images, even for the same galaxy at a given moment in time \citep[see also][]{Blaizot:2023}.

\section{\lya\ Properties of \sphinx\ Galaxies}
\label{sec:lya_props}

We begin by comparing \lya\ emission properties of our sample of \sphinx\ galaxies with observations. The angle-averaged \lya\ luminosity functions have already been discussed \citep{Garel:2021,Katz:2022}, while the line-of-sight equivalent was recently presented in Figure~31 of \cite{spdrv1}, finding very good agreement with observational constraints.

\begin{figure*}
\includegraphics{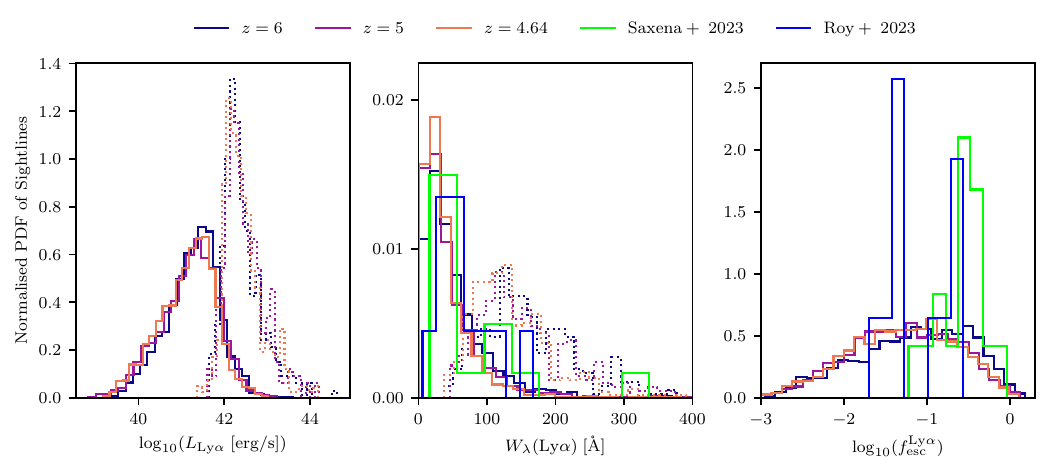}
    \caption{Distributions of \lya\ properties of our sample, including the \lya\ luminosity (\textit{left}), \lya\ equivalent width (\textit{centre}) and \lya\ escape fraction (\textit{right}). Solid lines are given by dust-attenuated quantities, while intrinsic values are given where possible by dotted lines. Comparisons are given to observational data from \protect\cite{Saxena:2023} and \protect\cite{Roy:2023} in green and blue respectively.}
    \label{fig:lya_pop_histograms}
\end{figure*}

Here we discuss the distributions of line-of-sight \lya\ luminosities, equivalent widths and escape fractions\footnote{Note that when viewed along individual sight lines, the Ly$\alpha$ escape fractions can be $>100\%$ due to resonant scattering of photons emitted away from an observers sight-line back into it \citep[see e.g.][]{Verhamme:2012}.}. Figure~\ref{fig:lya_pop_histograms} shows histograms of these quantities for the three redshifts considered compared to observational data spanning the redshift range $z \approx 3 - 8$ \citep{Saxena:2023, Roy:2023}. Where possible, both the intrinsic (dotted) and dust-attenuated (solid) values are shown. The effect of IGM attenuation on these properties is discussed in Appendix \ref{sec:igm}. We find little redshift dependence, apart from the fact that galaxies at $z = 6$ tend to exhibit slightly larger equivalent widths and \lya\ escape fractions. We also note that observational studies of LAEs are unlikely to include many galaxies with $f_{\rm esc}^{\rm Ly\alpha} \lesssim 1\%$. We see that dust attenuation decreases both the total \lya\ luminosity and equivalent widths. This behaviour is consistent with the picture outlined in Section 4.1 of \cite{Verhamme:2012}. Here, most UV continuum and \lya\ photons are emitted from the cores of dense, star-forming clouds. The UV continuum photons are then strongly extinguished before escaping these birth clouds, as modeled by \cite{Charlot:2000}. In contrast, the \lya\ photons are scattered many times within the birth clouds before having a chance to escape. This increases the distance travelled by them and thus attenuates them more significantly than the UV photons. 

\begin{figure}
    \includegraphics{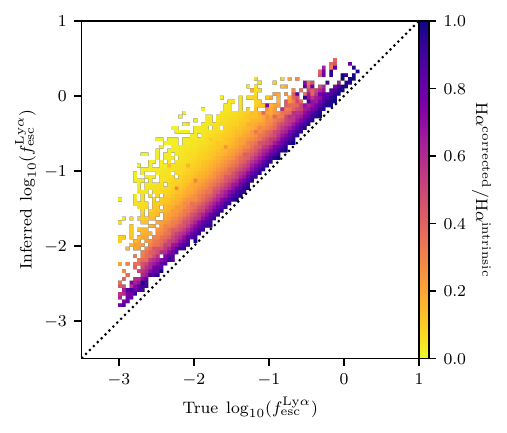}
    \caption{Histogram of Inferred \fesca\ (Equation \protect\ref{eq:lya_inf}) as a function of true \fesca\. The colour of each bin is given by the average ratio of dust-corrected to intrinsic H$\alpha$ contained therein. The one-to-one relation is shown as a dotted line. We find that dusty galaxies with insufficient flux correction tend to over-predict \fesca. We find a mean absolute error of $0.43$ dex}
    \label{fig:lyafesc_vs_lyafesc}
\end{figure}

\begin{figure*}
    \includegraphics{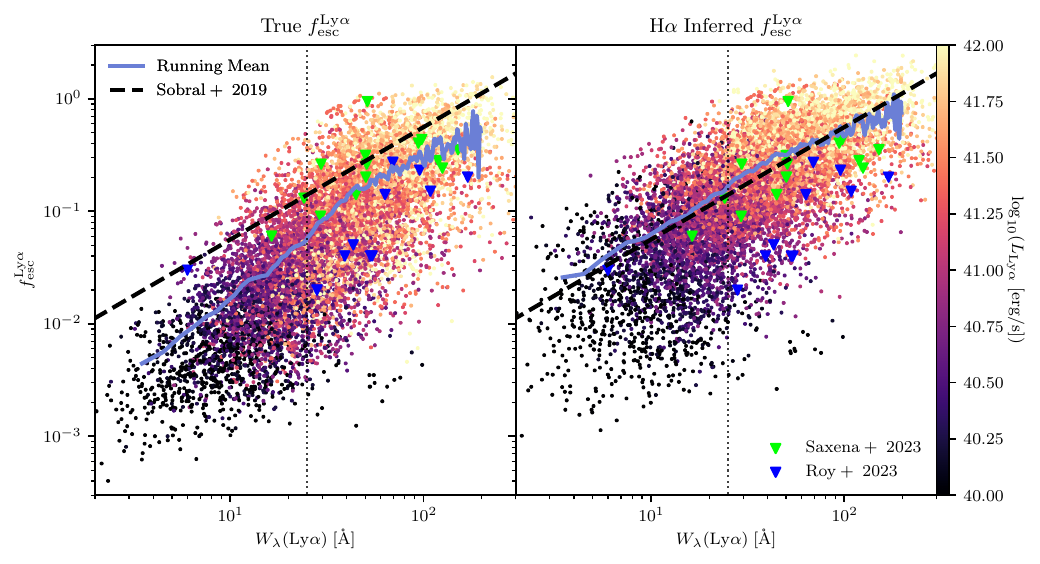}
    \caption{\lya\ escape fractions as a function of \lya\ equivalent widths for galaxies in \sphinx, coloured by their observed \lya\ luminosity. On the \textit{left}, we use the true \lya\ escape fraction as computed by {\small RASCAS}, while on the \textit{right} we use the H$\alpha$-inferred value as computed by Equation \ref{eq:lya_inf}. The empirical estimator for $2.2 \lesssim z \lesssim 2.6$ from \protect\cite{Sobral:2019} is shown in dashed black, along with our running mean in lilac. The typical observational cut for LAEs of $W_{\lambda}({\rm Ly\alpha}) \geq 25$\angstrom~ is also shown in black. Observational data from \protect\cite{Saxena:2023} and \protect\cite{Roy:2023} are given as a comparison.}
    \label{fig:lya_fesc_ew}
\end{figure*}
It is important to establish that the \lya\ escape fraction discussed in this work is different to that inferred observationally. While we compute \fesca\ by directly calculating the number of \lya\ photons to escape using {\small RASCAS}, traditionally this is done by inferring the intrinsic \lya\ flux from the ${\rm H\alpha}$ (or some other Balmer) line \citep[e.g.][]{Osterbrock:1989},
\eq{
f_{\rm esc}^{\rm Ly\alpha} = \frac{L_{\rm Ly\alpha}}{8.7 L_{\rm H\alpha} \times 10^{0.4A_{\rm H\alpha}}},
\label{eq:lya_inf}
}
where $L_{\rm Ly\alpha}$ and $L_{\rm H\alpha}$ are the dust-attenuated luminosity of \lya\ and H$\alpha$ respectively. The denominator is then the estimated intrinsic H$\alpha$ luminosity, inferred using the SMC dust attenuation law, $A_{\rm H\alpha}$ \citep{Gordon:2003}\footnote{This was done using the {\small dust-extinction PYTHON} package (v1.2) available at \url{https://github.com/karllark/dust_extinction}. The correction for H$\alpha$ is then $A_{\rm H\alpha} = 2.17 {\rm E(B - V)}$, where $\rm E(B - V)$ is measured using the Balmer decrement for each line of sight.}, converted to the intrinsic \lya\ luminosity by means of a canonical factor discussed below. Figure~\ref{fig:lyafesc_vs_lyafesc} shows a histogram of the true and inferred (from Equation~\ref{eq:lya_inf}) \fesca\, where each bin is coloured by the average ratio of attenuation-corrected to true intrinsic ${\rm H\alpha}$ luminosity. We find a mean absolute error between the two methods of $0.43$ dex. Here, we see very clearly that this method of inferring \fesca\ \citep[which is typically used in observational studies, e.g.][]{Hayes:2005,Sobral:2019, Flury:2022b,Saxena:2023,Roy:2023} is only close to accurate when the dust-correction works sufficiently well to recover the intrinsic emission (i.e. for the purple pixels). To understand this, it is important to consider that Equation \ref{eq:lya_inf} makes use of several key assumptions, which clearly break down for the orange pixels. First, it uses the fact that under case B assumptions, the intrinsic Ly$\alpha$-to-H$\alpha$ ratio is fixed at some value (here, 8.7). In reality, this value depends on the electron number density as well as particularly the electron temperature (typically $n_e \sim \SI{300}{\per\centi\metre\cubed}$ and $T_e\sim\SI{1e4}{\kelvin}$ are used). These are also assumed to be constant for a given galaxy, whereas fluctuations in such properties across galaxies are known to bias integrated measurements \citep[e.g.][]{Cameron:2023}. Next, this analysis ignores contributions from the collisional emissivity to \lya\ emission. While subdominant, this contribution can be important \citep[see e.g.][for discussions]{Kimm:2019,Mitchell:2021,Smith2022b}. For example, Figure 8 of \cite{Kimm:2019} suggests that collisional radiation contributes about $\sim25\%$ of the intrinsic \lya\ emission at the giant molecular cloud scale. This would imply an upper-limit systematic shift of 0.1 dex which would help to explain the systematic over-estimate in Figure \ref{fig:lyafesc_vs_lyafesc}. Moreover, it is important to note that the collisional component is less affected by dust than the recombination component. This is due to the fact that while recombination photons are predominantly produced in dense, cold gas \citep[as discussed above, see also][]{Charlot:2000}, collisional photons are also produced significantly in more diffuse gas where it is easier for them to escape unencumbered. Finally, the geometry of the dust-screen model is important. Methods such as Equation \ref{eq:lya_inf} typically assume a uniform dust screen. However, it has been shown that clumpy dust screen models better reproduce theoretical line ratios \citep[see][and references therein]{Scarlata:2009}, while also representing the more realistic and complex dust geometry found in simulated galaxies. 

Beyond including the modelling of the contributions discussed above, another strategy to better infer \fesca\ at lower redshifts may be to make use of detections of other transitions of hydrogen -- most notably the Paschen series. For example, recently \cite{Reddy:2023} used JWST/NIRSpec observations of Paschen lines in galaxies at $1 \lesssim z \lesssim 3$ to re-evaluate dust-extinction curves, finding that Balmer-inferred estimates were insufficient. While this evidence is marginal, it certainly points to the need to better understand dust attenuation and dust-obscured regions. Another option would be to exploit the synergy of rest-frame UV/NIR JWST observations of galaxies during the epoch of reionization with rest-frame IR/FIR measurements from ALMA, a technique which is beginning to be explored \citep[e.g.][]{Rujopakarn:2023}. Here, the total dust emission can be used to better estimate the attenuation of UV/optical emission lines. However, at such redshifts the point-spread function and sensitivity of the instrument can rapidly become an issue.

This over-estimate is likely to be unimportant in studies comparing \fesca\ to the escape of ionizing radiation, as sight-lines for which the dust-correction works well are likely to have little dust and therefore significant LyC leakage. Otherwise, this bias is clearly important to include in error estimation of reported \lya\ escape fractions.

Next, we explore the effect of this over-estimate in \fesca on the $f_{\rm esc}^{\rm Ly\alpha}-W_{\lambda}({\rm Ly\alpha})$ relation. Figure \ref{fig:lya_fesc_ew} shows \fesca\ as a function of \lya\ equivalent widths, coloured by the observed \lya\ luminosity, compared to high-redshift observations \citep{Saxena:2023,Roy:2023}. The empirical estimator from \cite{Sobral:2019} for $2.2 \lesssim z \lesssim 2.6$ is plotted in dashed black, as well as the running mean for \sphinx\ in lilac. On the \textit{left}, we show the true \fesca, while on the \textit{right} we use the H$\alpha$-inferred \fesca\ as given by Equation \ref{eq:lya_inf}. We find that these three quantities are well correlated. However, we find that comparisons with the relation from \cite{Sobral:2019} depend strongly on the version of \fesca\ used. Specifically, this estimator tends to systematically over-estimate the true \lya\ escape fractions of our mock observations by $\sim0.3$~dex. As discussed above, in the case of the H$\alpha$-inferred values however, \lya\ escape fractions become over-estimated, shifting the distribution closer to the relation from \cite{Sobral:2019}. Furthermore, lines-of-sight that produce lower \lya\ equivalent widths and luminosities (and therefore tend to be dustier) tend to more severely under-predict the intrinsic luminosity of H$\alpha$, therefore over-predicting \fesca\ even more. This has the effect of reducing the gradient of the running mean, such that it matches the empirical estimator remarkably well. Overall, our mock observations of \sphinx\ galaxies reproduce the trends and spread from JWST observations of \lya\ emission in the high-redshift Universe, as shown by the points in blue and green.

In general, the numerator of the \lya\ equivalent width is set by the \lya\ escape fraction, implying that we should expect the correlation evident in Figure \ref{fig:lya_fesc_ew}. However, \wlya\ also depends on the impact of sSFR and local ISM properties on the continuum at 1216~\AA. Therefore, one would expect a correlation between the two quantities. However, matching the observational trend perfectly also requires the local state of the ISM to be realistic in order to produce the correct intrinsic emission \citep{Osterbrock:1989}. Therefore, we can be cautiously optimistic that the multi-phase ISM of \sphinx\ reproduces conditions in and around real H~{\small II} regions reasonably well.

\section{\lya\ Diagnostics for the LyC Escape Fraction}
\label{sec:results}

\begin{figure*}
    \includegraphics{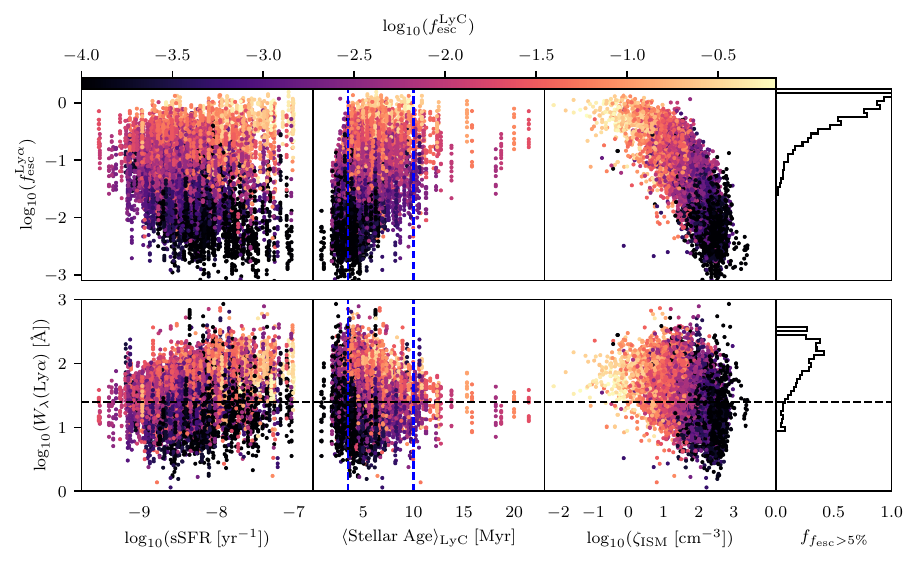}
    \caption{(\textit{Top}) \lya\ escape fraction, \fesca, as a function of sSFR (\textit{left}), LyC luminosity-weighted mean stellar age (\textit{centre}) and neutral gas attenuation parameter $\zeta_{\rm ISM}$ (\textit{right}) coloured by the global LyC escape fraction, \fesc. The attached histogram (right) shows the fraction of galaxies for a given \fesca\ with $f_{\rm esc}^{\rm LyC} > 5\%$ for each bin that contains at least 5 such galaxies. While not correlating well with sSFR, high \fesca\ selects for galaxies with mean stellar ages $> \SI{3.5}{\mega\year}$ and low $\zeta_{\rm ISM}$. Therefore, \fesca\ is a good indicator for LyC escape. (\textit{Bottom}) Same as above, but for \lya\ equivalent widths. \wlya\ correlates weakly with sSFR but does not trace the mean stellar age. However, larger \lya\ equivalent widths select for galaxies with lower $\zeta_{\rm ISM}$. Thus, \wlya\ is a necessary but insufficient diagnostic for \fesc.}
    \label{fig:big_crit_1}
\end{figure*}

We are now able to test the efficacy of known \lya\ diagnostics for LyC escape from \sphinx\ galaxies. In order to understand the underlying physics, we utilise the framework presented in \cite{Choustikov:23}, which argues that a good diagnostic for high LyC leakage should:
\begin{itemize}
    \item select for galaxies with a high specific star formation rate (sSFR),
    \item be sensitive to stellar population ages $\geq \SI{3.5}{\mega\year}$ and $\leq \SI{10}{\mega\year}$ to catch galaxies that have undergone supernovae feedback, but are still producing ionizing radiation,
    \item contain a proxy for the density and neutral state of the ISM.
\end{itemize}

Here, we define sSFR as the 10~${\rm Myr}$-averaged star formation rate normalised by the stellar mass, we use a mean stellar population age weighted by the ionizing luminosity contribution of each star particle (in order to focus on the population crucial to producing LyC photons) and, following \cite{Choustikov:23}, define the composite parameter $\zeta_{\rm ISM} = {\rm E(B - V)}\times \langle n_{\rm HI}\rangle_{[\rm OII]} / [{\rm cm}^{-3}]$ as the product of the UV attenuation as well as the $[{\rm OII}]\; \lambda\lambda 3726,3728$-weighted density of neutral hydrogen. This is used for demonstrative purposes to isolate information about the density and neutral state of the ISM. This proxy is used as it is difficult to find a suitable property which is uncorrelated with the age or SFR of the galaxy. It is important to note that galaxies falling outside of these criteria are \textbf{not} omitted from any analysis or discussion in this work. We use this framework to understand correlations (or lack thereof) between \lya\ properties and LyC leakage.

When comparing with direct observational data, we use the line-of-sight values for the LyC escape fraction (\fesclos). Our goal however is to understand correlations with the global (i.e. angle-averaged) LyC escape fraction (\fesc) as this is the relevant quantity for reionization. Scatter between these two quantities is discussed in Appendix \ref{sec:aa_los}. All \lya\ quantities discussed are dust-attenuated using {\small RASCAS} and observed along a given sight-line.

\subsection{\lya\ Escape Fraction, \fesca}
\label{sec:fesca}

It is well-established that the \lya\ and LyC escape fractions are well-correlated \citep[e.g.][]{Dijkstra:2016, Verhamme:2017, Steidel:2018,Gazagnes:2020,Izotov:2021, Pahl:2021, Katz:2022, Maji:2022}. This is expected, given the fact that both of these photons can be absorbed by similar components of the ISM. However, LyC photons can be absorbed by both dust and H~{\small I} while \lya\ is mainly absorbed by dust\footnote{Deuterium and molecular hydrogen can also absorb \lya\ photons, but their effect is minimal.} and can be scattered back into a given sight-line. As a result, is clear that one would naively expect \fesca\ to be greater than \fesc for a given sight-line.

In the top row of Figure \ref{fig:big_crit_1}, we plot the \lya\ escape fraction as a function of sSFR (\textit{left}), mean stellar population age (\textit{centre}) and the composite ISM parameter $\zeta_{\rm ISM}$ (\textit{right}), coloured by \fesc. While \fesca\ does not correlate with sSFR, we know from Figure~\ref{fig:lya_fesc_ew} that it correlates with $L_{\rm Ly\alpha}$, which itself does correlate with SFR \citep[e.g.][]{Sobral:2018}. As a result, selecting for galaxies with very high \fesca\ does marginally systematically select for systems with higher sSFR. Next, we find that observations with high \fesca\ match very well with galaxies within the correct stellar population age, particularly selecting against galaxies with ages $\lesssim \SI{3.5}{\mega\year}$ where the birth clouds are yet to be disrupted by SNe.  Finally, as expected we find that \fesca\ correlates with the state of the ISM. Lines-of-sight with less neutral hydrogen and dust attenuation tend to produce higher \lya\ escape fractions \citep[e.g.][]{Verhamme:2015,Verhamme:2017, Dijkstra:2016, Dijkstra:2017}. As a result, \fesca\ selects for galaxies with the optimal mean stellar ages and traces $\zeta_{\rm ISM}$, thus satisfying two criteria very well, making it a good indicator for high LyC escape fractions. This is seen particularly easily in the histogram (\textit{top-right} of Figure~\ref{fig:big_crit_1}), which shows the fraction of galaxies in each bin of \fesca\ which have $f_{\rm esc}^{\rm LyC} \geq 5\%$.

\begin{figure*}
    \includegraphics{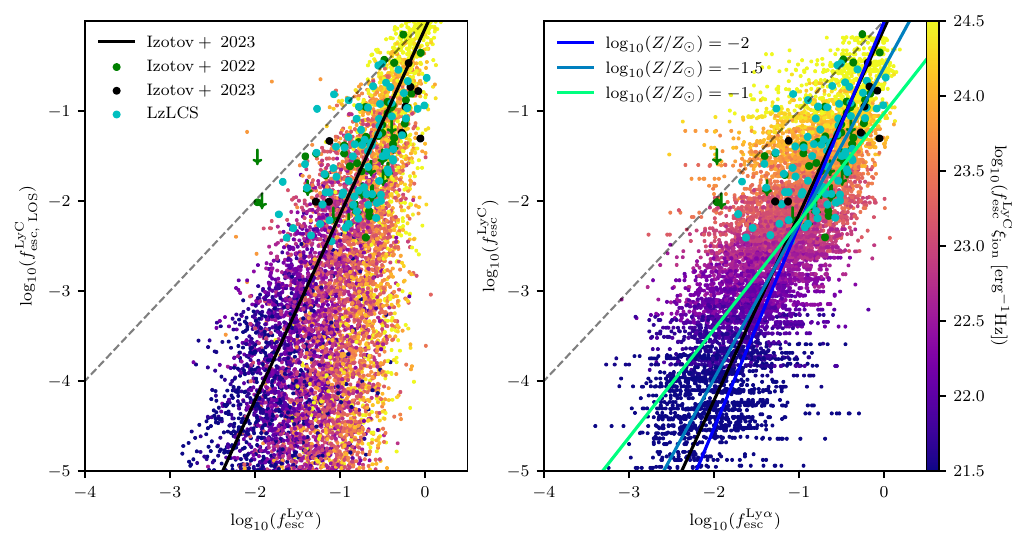}
    \caption{(\textit{Left}) Line-of-sight LyC escape fraction versus line-of-sight \lya\ escape fraction, coloured by \fx. The one-to-one relation is shown as a dashed line. We over-plot observational data from \protect\cite{Flury:2022b}, \protect\cite{Izotov:2022}, and \protect\cite{Izotov:2023} in cyan, green, and black, respectively. (\textit{Right}) Global LyC escape fraction as a function of line-of-sight \fesca, coloured by \fx. We show the one-to-one relation, the relation given by \protect\cite{Izotov:2023}, as well as lines of best fit given by Equation \protect\ref{eq:fesca} for $\log_{10}(Z/Z_{\odot}) \in \{-2, -1.5, -1\}$. In both cases, we find a good correlation between the two quantities. A clear gradient is visible. Galaxies with larger \fesca\ effectively contribute more to reionization of the IGM, as they have the largest global values of \fx.}
    \label{fig:fesc_fesca}
\end{figure*}

Figure~\ref{fig:fesc_fesca} shows both variations in the line-of-sight LyC escape fraction, \fesclos\ (\textit{left}) as well as the global LyC escape fraction, \fesc\ (\textit{right}) as a function of the \lya\ escape fraction, coloured by \fx. We also include observational data \citep{Izotov:2022,Izotov:2023,Flury:2022b} in green, black, and cyan respectively, as well as the line of best fit from \cite{Izotov:2023} in black. We find that our mock observations reproduce the observational trend very well. Indeed, as discussed above it is clear that \fesca\ is a suitable diagnostic for the LyC escape fraction, with \fesca\ being a somewhat stronger predictor of the global than line-of-sight LyC escape fraction. Furthermore, we find that in general, $f_{\rm esc}^{\rm LyC}  < f_{\rm esc}^{\rm Ly\alpha}$, with the systems for which this is not the case having the largest \fx. In both cases we find that mock observations with larger \lya\ escape fraction tend to have larger \fx, suggesting that galaxies with high \lya\ escape fractions might contribute decisively to reionization. However, it is clear from the \textit{right} panel that there is significant scatter in this relation. 

This scatter is driven predominantly by the line-of-sight-dependent nature of \lya, where \lya\ photons can scatter out of or into a given aperture. However, it is important to note that the sources of \lya\ (gas) and LyC (mostly stars) photons are distributed differently in galaxies, and therefore have different escape channels. For example, \cite{Maji:2022} found that six of their analysed {\small SPHINX} galaxies had $f_{\rm esc}^{\rm LyC} > f_{\rm esc}^{\rm Ly\alpha}$, corresponding to galaxies with dusty, low H~{\small I} density escape channels close to their centres. Indeed, in Figure \ref{fig:fesc_fesca} we find a number of such systems too, particularly when global \fesc\ is considered, agreeing too with low-redshift observations .

There is a built-in dependence for the global LyC escape fraction on metallicity due to our implementation of dust. Namely, in systems with little metallicity and dust, the \lya\ escape fraction will be close to 100\%, irrespective of the LyC escape fraction (which can be orders of magnitude lower). To this end, we include a best-fit relation depending on both the \lya\ escape fraction and metallicity given by:

\eq{
\log_{10}(f_{\rm esc}^{\rm LyC}) = \ls 0.1429 - 1.057\log_{10}(Z/Z_{\odot}) \rs\log_{10}(f_{\rm esc}^{\rm Ly\alpha}) \nonumber\\
 -1.015\log_{10}(Z/Z_{\odot}) - 2.014, \label{eq:fesca}
}

Lines are over-plotted for $\log_{10}(Z/Z_{\odot}) \in \{-2, -1.5, -1\}$. Here, we can see that the gradient of this relation is most strongly affected, steepening for metal-poor populations with less dust, in line with the above discussion. As expected, the line for low-intermediate metallicities of $\sim0.03Z_{\odot}$ visually fits our data best.

\subsection{\lya\ Equivalent Width, \wlya}
\label{sec:ew}

Previously, the \lya\ equivalent width, \wlya\ has been found to depend strongly on the H~{\small I} covering fraction, as well as on the optical depth \citep[e.g.][]{Reddy:2016,Gazagnes:2018}. As a result, it is reasonable to expect that greater \wlya\ should correlate with increased \fesc. It has also been found that \lya\ doublets with large red to blue ratios tend to have higher equivalent widths \citep{Blaizot:2023}. Given the fact that such signals tend to correspond to outflows (signatures of effective feedback), it further suggests that this trend should exist provided that LyC leakage is a feedback-regulated quantity \citep[e.g.][]{Trebitsch:2017}. \cite{Steidel:2018} and \cite{Pahl:2021} found a very strong correlation between \fesc\ and \wlya, while a slightly weaker correlation was found by the LzLCS \citep{Flury:2022b}, for the Green Peas \citep[GPs,][]{Henry:2015,Yang:2017}, and galaxies from the VANDELS survey \citep{Saldana-Lopez:2023}. 

Figure~\ref{fig:big_crit_1} (\textit{bottom row}) shows \wlya\ as a function of sSFR, mean stellar age and $\zeta_{\rm ISM}$ coloured by \fesc. We find that \wlya\ correlates very weakly with sSFR. Furthermore, greater equivalent widths tend to select for galaxies with stellar ages $\lesssim \SI{10}{\mega\year}$, while selecting for systems with less neutral gas and UV attenuation. As a result, \wlya\ satisfies two criteria weakly, making it a helpful but insufficient diagnostic for LyC leakage. We do note however that the typical selection cut of equivalent widths greater than 25~\AA~ produces a sample with a much larger number of strong LyC leaking galaxies, as seen in the \lya\ histogram (\textit{bottom-right}) of Figure~\ref{fig:big_crit_1}. We also see a slight decrease in the fraction of galaxies with $f_{\rm esc}^{\rm LyC} > 5\%$ in bins with equivalent widths of $W_{\lambda}({\rm Ly\alpha}) \gtrsim 250$~\AA. This is due to the fact that such a sample is contaminated by galaxies with young stellar populations ($\lesssim \SI{3.5}{\mega\year}$) and large $\zeta_{\rm ISM}$ that have yet to clear the dusty ISM surrounding the stellar nursery in order to let LyC photons to escape. 

\begin{figure}
    \includegraphics{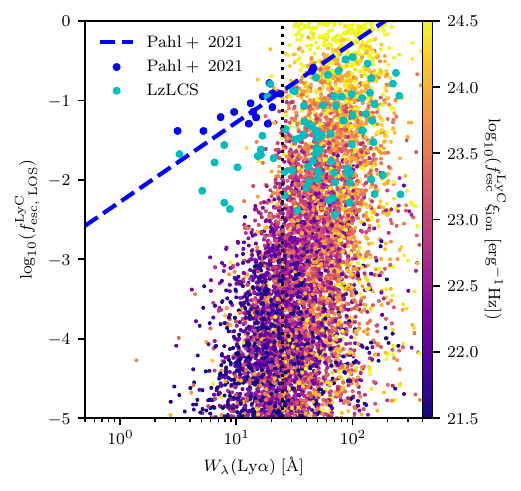}
    \caption{Line-of-sight LyC escape fraction as a function of \lya\ equivalent widths, coloured by \fx. This is compared to observational data from \protect\cite{Pahl:2021}, and \protect\cite{Flury:2022b} in blue and cyan respectively. The relation from \protect\cite{Pahl:2021} is also shown as a dashed blue line. We find that observations with large \fesclos\ all have $W_{\lambda}({\rm Ly\alpha}) > 25$~\AA, and are therefore bright LAEs. This however is not definitively indicative of LyC leakage.}
    \label{fig:fesc_ew}
\end{figure}

Figure~\ref{fig:fesc_ew} shows the line-of-sight LyC escape fraction as a function of \wlya\ coloured by \fx, compared to observational data \citep{Flury:2022b,Pahl:2021} in cyan and blue respectively. We find some correlation (albeit with a lot of scatter) between the two quantities, with virtually all strong leakers being bright LAEs ($W_{\lambda} > 25$\angstrom). While we find some good agreement with \cite{Flury:2022b}, we find poor agreement with the rest of the observational data as well as the relation from \cite{Pahl:2021}, particularly for bright LAEs. This may be due to selection effects as the galaxies used in stacks by \cite{Steidel:2018} and \cite{Pahl:2021} are significantly more UV-bright than those generally found in \sphinx, filling out the region of galaxies with moderate \fesclos\ and low $W_{\lambda}({\rm Ly\alpha})$. We find that the \sphinx\ galaxies that best compare to these stacks tend to have the strongest UV flux and youngest mean stellar population ages, corresponding to starburst galaxies that are yet to experience strong feedback.

There is another effect to consider. If the LyC escape fraction is close to 100\%, then the neutral hydrogen column density must be very low. As a result, this case should produce lower \lya\ equivalent widths \citep[e.g.][]{Nakajima:2014, Steidel:2018} as fewer \lya\ photons are produced. However, \lya\ equivalent widths also depend strongly on the stellar continuum flux and stellar population source model. In reality, it is unclear whether we should expect such a bimodal distribution.

Interestingly, the performance of \wlya\ as an indicator for LyC leakage is similar to that of the equivalent width of Balmer lines such as H$\beta$ ($W_{\lambda}({\rm H\beta})$). Comparing this discussion with that of $W_{\lambda}({\rm H\beta})$ in a previous study (see Section 4.1.5 of \citealt{Choustikov:23}), we find that larger \lya\ equivalent widths correlate worse with sSFR but are less polluted with galaxies with mean stellar ages younger than $\SI{3.5}{\mega\year}$. We conclude that the equivalent widths of Hydrogen lines should only be used in combined diagnostics for LyC escape fractions, such as the $W_{\lambda}({\rm H\beta})-\beta$ diagram \citep{Zackrisson:2013, Zackrisson:2017}. In general, we find that large \lya\ equivalent widths are a necessary but insufficient criterion for strong LyC leakage. Selecting for bright LAEs (as indicated by the vertical dotted line in Figure \ref{fig:fesc_ew}) manages to capture all strong line-of-sight LyC leakers. Therefore, \wlya\ can be used as a useful first stage in any attempt to find LyC leakers from a sample of galaxies with confirmed \lya\ emission.

\subsection{Peak Separation, \vsep}
\label{sec:vsep}

\lya\ photons must scatter away from line center in order to escape. As a result, this process often drives the production of a double peak in \lya\ spectra. In such cases, the velocity shift corresponding to the separation between the two peaks of the doublet, \vsep\ is a natural quantity to investigate. Due to the fact that scattering in frequency space depends on the H~{\small I} column density, from a theoretical basis $N_{\rm HI}$ should correlate with \vsep\ \citep[e.g.][]{Dijkstra:2006,Verhamme:2015, Kakiichi:2021} and therefore we expect \vsep\ to correlate with \fesc \citep[see also ][]{Kimm:2019,Yuan:2024}. Observationally, the relationship between \vsep\ and \fesc\ has been well explored \citep[e.g.][]{Verhamme:2017,Izotov:2018b, Flury:2022b}. 

\begin{figure*}
    \includegraphics{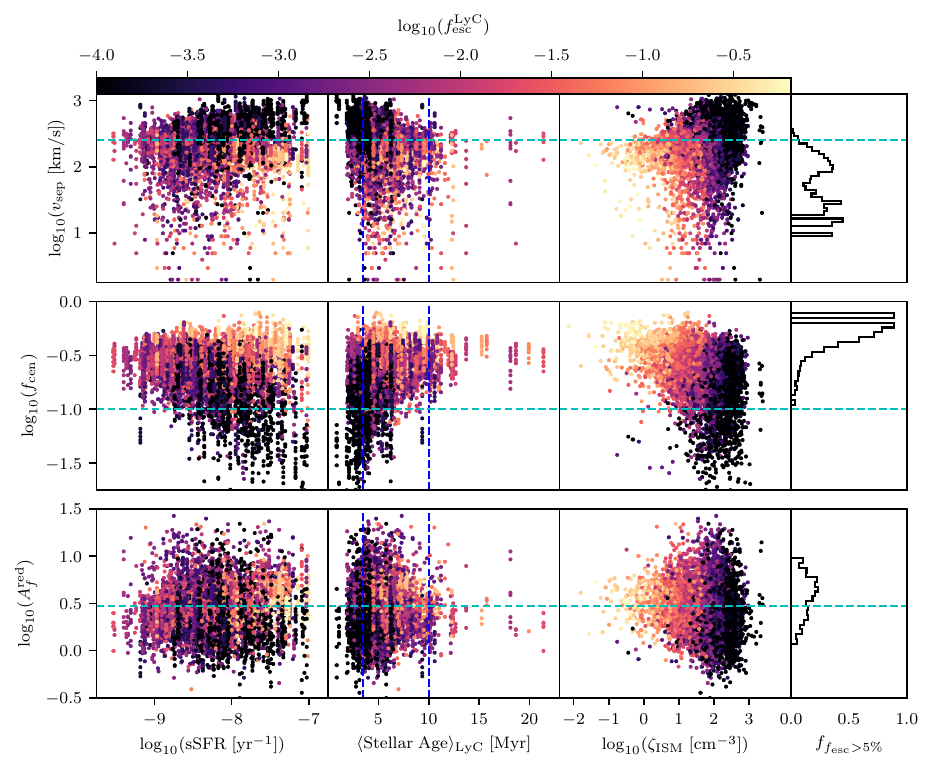}
    \caption{(\textit{Top}) As in Figure \ref{fig:big_crit_1}, but for the \lya\ peak separation. Mock observations with $v_{\rm sep} > \SI{250}{\kilo\metre\per\second}$ (cyan line) select for galaxies with young stellar populations and large $\zeta_{\rm ISM}$. Therefore, low peak separations are found in galaxies with higher LyC escape fractions. (\textit{Middle}) As above, but for the central escape fraction. We find that large values of \fcen\ (compared to the $10\%$ cut suggested by \protect\cite{Naidu:2022} in cyan) select for galaxies with high sSFR, stellar ages in the range required and low $\zeta_{\rm ISM}$. Thus, for large values of \fcen\ it is a suitable diagnostic for the LyC escape fraction. (\textit{Bottom}) As above, but for the red peak asymmetry. We find that \asym\ does not trace sSFR or the mean stellar age. However, galaxies with the lowest $\zeta_{\rm ISM}$ tend to be clustered around $A_{f}^{\rm red} \sim 3$ (shown in cyan). Thus, \asym\ is not by itself a good indicator for the presence of LyC escape, though it can be used to infer the method of escape \protect\citep{Kakiichi:2021}.}
    \label{fig:big_crit_2}
\end{figure*}

The \textit{top row} of Figure~\ref{fig:big_crit_2} shows \vsep\ as a function of sSFR, mean stellar age, and $\zeta_{\rm ISM}$ colored by the global \fesc. We find no trend between \vsep\ and sSFR. Next, we find that spectra with large peak separations ($> \SI{250}{\kilo\metre\per\second}$) select for younger mean stellar populations with more UV attenuation and a higher neutral gas density. Therefore, these systems are unlikely to be LyC leakers, agreeing with previous work \citep[e.g.][]{Verhamme:2015,Naidu:2022, Flury:2022b}. As a result, we find that \lya\ peak separations $\lesssim \SI{250}{\kilo\metre\per\second}$ weakly fulfill two of the criteria, making it a potentially useful but insufficient diagnostic for LyC leakage.

\begin{figure}
    \includegraphics{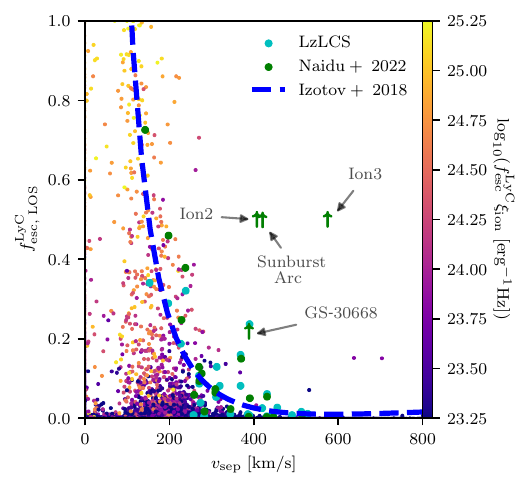}
    \caption{Line-of-sight LyC escape fraction as a function of \lya\ peak separation (for all cases where there are two peaks in the \lya\ spectrum) coloured by \fx. We include observational data from \protect\cite{Flury:2022b} and \protect\cite{Naidu:2022} in cyan and green respectively, as well as the relation from \protect\cite{Izotov:2018b} as a dashed blue line. We find that mock observations with $v_{\rm sep} < \SI{250}{\kilo\metre\per\second}$ are much more likely to be significant LyC leakers, but this is not a sufficient condition.}
    \label{fig:fesc_vsep}
\end{figure}

Figure \ref{fig:fesc_vsep} shows \fesclos\ versus \vsep\footnote{\lya\ spectra with single peaks are assigned a \vsep\ of zero so that they are not omitted.}\ for our mock observations as a function of \fx. In order to compare to observations, we over-plot data from the LzLCS \citep{Flury:2022b} and \citet{Naidu:2022} in cyan and green, respectively, as well as the relation from \cite{Izotov:2018b} in blue. We find some agreement (with a large amount of scatter), along with the fact that the majority of strong leakers have $v_{\rm sep} < \SI{250}{\kilo\metre\per\second}$. However, there is no strong trend present. Figure \ref{fig:fesc_vsep} also indicates the existence of an envelope, suggesting that it is very unlikely to find galaxies with large \vsep\ and high \fesc. Furthermore, there is a perhaps unexpectedly large population of galaxies with low \vsep\ and low \fesc. This, along with the fact that our mock observations tend to have systematically lower peak separations is likely due to the fact that \sphinx\ galaxies are altogether less massive and less UV-bright than our comparative sample. This is discussed further in Appendix~\ref{sec:homog}, but suggests that selection effects may have a role to play.

\subsection{Central Escape Fraction, \fcen}

Recently, \cite{Naidu:2022} proposed a new \lya\ indicator for LyC escape, $f_{\rm cen}$. This represents the ratio of \lya\ flux within $\pm \SI{100}{\kilo\metre\per\second}$ of line centre to the total \lya\ flux\footnote{Specifically, we follow \cite{Naidu:2022} and use the excess flux within a window of $\pm \SI{1000}{\kilo\metre\per\second}$ to capture the total flux.}. This is crucial, as line-centre emission\footnote{Due to the fact that each galaxy has a peculiar velocity, we assume that H$\alpha$ provides the true redshift and shift the spectra accordingly.} (corresponding to large values of $f_{\rm cen}$) may be indicative of \lya\ photons that escape at line center with little scattering, thus tracing low-column density channels \citep{Behrens:2014b}. These may also be optically thin to LyC photons \citep[e.g.][]{Harrington:1973,Neufeld:1991}, thus suggesting that the two quantities should be well correlated. 

As before, the \textit{middle row} of Figure~\ref{fig:big_crit_2} shows \fcen\ as a function of sSFR, mean stellar age and $\zeta_{\rm ISM}$. For clarity, we also include the cut suggested by \cite{Naidu:2022} in cyan. We find no strong trend between \fcen\ and sSFR, beyond that empirically, lines-of-sight with $f_{\rm cen} \gtrsim 50\%$ tend to have systematically higher sSFRs. Next, we find no strong relationship between \fcen\ and mean stellar age, apart from the fact that lines-of-sight with $f_{\rm cen} < 10\%$ tend to have young stellar populations and lines-of-sight with $f_{\rm cen} \gtrsim 40\%$ tend to have stellar populations in the correct age range (indicated by vertical blue lines). As a result, we see that observations with $f_{\rm cen} > 10\%$ are significantly more likely to have LyC leakage, while observations with $f_{\rm cen} \gtrapprox 40\%$ are overwhelmingly likely to be strong LyC leakers. Finally, galaxies with larger \fcen\ tend to correspond to sight lines with less dust attenuation and neutral gas. Combining these facts, we find that the central escape fraction marginally satisfies all three criteria when $f_{\rm cen}\gtrsim 40\%$, making it a potentially good diagnostic for LyC leakage for this range of values, when also combined with other information. This is also seen in the histogram for \fcen\ (\textit{middle-right}), where bins of larger \fcen\ tend to have a higher fraction of galaxies with $f_{\rm esc}^{\rm LyC} > 5\%$.

\begin{figure}
    \includegraphics{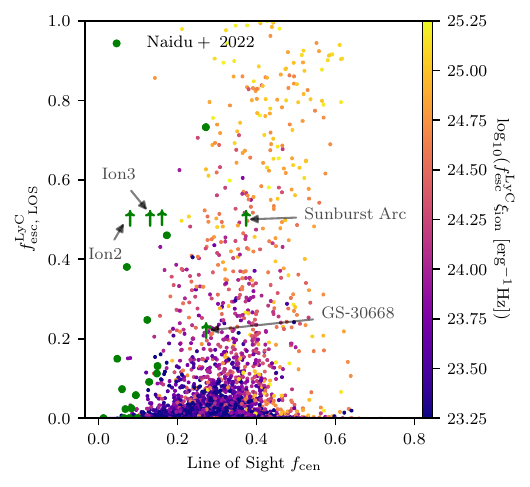}
    \caption{Line-of-sight LyC escape fraction as a function of central \lya\ escape fraction, coloured by \fx. We include observational data from \protect\cite{Naidu:2022} in cyan. We find that in general, galaxies with larger \fcen\ have higher LyC escape fractions, as well as contributing more to reionization (by having a larger value of \fx). Particularly, we find that contamination from non-leaking galaxies drops off after $f_{\rm cen} \gtrsim 0.4$.}
    \label{fig:fesc_fcen}
\end{figure}

Figure \ref{fig:fesc_fcen} shows \fesclos\ as a function of \fcen\ for our mock observations coloured by \fx. We include observational data from \cite{Naidu:2022} in green for comparison. We find a large amount of scatter between the two quantities, however galaxies with larger \fcen\ tend to leak more LyC photons. Interestingly, we find in practise that there is a marginally stronger correlation between \fcen\ and \fesc\ than that with \fesclos. Furthermore, while the cut of $f_{\rm cen} > 10\%$ discussed above captures the vast majority of strong leakers, we find that this is an insufficient diagnostic as the sample is highly impure (see discussion in Section \ref{sec:using_lya}). We note that the fact that our average \fcen\ values are skewed systematically larger than those reported by \cite{Naidu:2022}. This offset is also likely due to the fact that \sphinx\ galaxies are less massive and have smaller H~{\small I} masses than the comparative observational sample. In fact, this offset is also consistent with the offset found in \vsep, as discussed further in Appendix \ref{sec:homog}. As a result, while \fcen\ has the potential to be a useful diagnostic for LyC leakage, we caution that there is likely a hidden H~{\small I} mass dependence that needs to be accounted for.

\subsection{Red Peak Asymmetry, \asym}

\begin{figure}
    \includegraphics{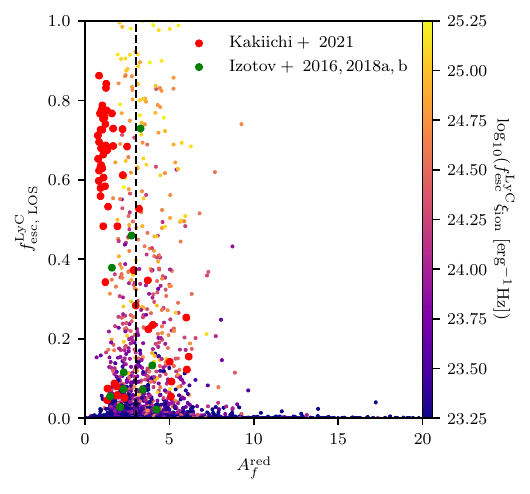}
    \caption{Line-of-sight LyC escape fraction as a function of \lya\ red peak asymmetry, coloured by \fx. We include observational data from \protect\cite{Izotov:2016, Izotov:2018b, Izotov:2018} in green as well as data from simulations in red \protect\citep{Kakiichi:2021}. We find that galaxies with significant LyC leakage tend to cluster around $A_f^{\rm red} \sim 3$, but that such a sample is contaminated by non-leakers. The strongest LyC leakers have larger values of \asym\ as compared to those from simulated giant molecular clouds. This is due to effects introduced by later reprocessing of \lya\ photons in the CGM \protect\citep{Blaizot:2023}.}
    \label{fig:fesc_A}
\end{figure}

Recently, \cite{Kakiichi:2021} suggested the red peak asymmetry as a diagnostic of LyC leakage. This asymmetry can be quantified as \asym,
\eq{
A_f^{\rm red} = \frac{\int_{v_{\rm red}}^{\infty}f_{\lambda}\; d\lambda}{\int_{v_{\rm valley}}^{v_{\rm red}}f_{\lambda}\; d\lambda},
}
where $v_{\rm red}$ and $v_{\rm valley}$ are the velocity shifts of the red peak and central valley respectively.

Using RHD simulations of giant molecular clouds with turbulence and radiative feedback, \cite{Kakiichi:2021} showed that in a turbulent H~{\small II} region, the flux of ionising photons from the central source will produce a mix of density- and ionisation-bounded channels, based on the low or high column density of neutral hydrogen respectively. LyC photons traversing the ionisation-bounded medium will then have a much larger probability of escape (i.e. contributing a larger \fesc) as compared to those passing through a density-bounded region. From the perspective of \lya\ photons, these two types of channels represent different types of escape \citep[e.g.][]{Gronke:2016,Gronke:2017}. In the ionisation-bounded (or high $N_{\rm HI}$) case, the gas remains optically thick far into the wings of the line. As a result, \lya\ photons must scatter multiple times in order to diffuse sufficiently far in frequency space as to escape. In contrast, in the density-bounded (or low $N_{\rm HI}$) case, the gas is optically thick around line center, becoming optically thin in the Lorentzian wings. Therefore, \lya\ photons are able to escape after a single interaction only, having been frequency-shifted sufficiently far from the Doppler core \citep[e.g.][]{Dijkstra:2017}. Here, \cite{Kakiichi:2021} found that systems with either extreme case of leakage produced \lya\ signals with low red peak asymmetry ($A_{f}^{\rm red} \lesssim 3$) with a mix of small and large \fesc. In contrast, configurations with a mixture of the two cases (i.e. with low column density `holes') produced larger red peak asymmetries ($A_{f}^{\rm red} \gtrsim 3$) with a mixture of low to intermediate LyC escape fractions. 

The \textit{bottom} row of Figure~\ref{fig:big_crit_2} gives \asym\ as a function of sSFR, mean stellar age and $\zeta_{\rm ISM}$, coloured by \fesc. The turnover value of $A_{f}^{\rm red} = 3$ from \cite{Kakiichi:2021} is highlighted in cyan. We find no significant dependence on sSFR or mean stellar age, but note that sight-lines with lower $\zeta_{\rm ISM}$ tend to be within the region of $A_f^{\rm} \sim 2-5$. Furthermore, the ISM resolution in \sphinx\ is insufficient to capture the escape of photons through small channels created by turbulence. This explains the drop-off in \fesc\ for $A_f^{\rm red} \gtrsim 6$ observed in Figure \ref{fig:fesc_A}. We note that while our results show some strong discrepancies with those of \cite{Kakiichi:2021}, one significant difference is the fact that \sphinx\ galaxies have a CGM. This extra stage of reprocessing of \lya\ photons has been shown to modify \lya\ signals, notably by increasing the asymmetry of red peaks \citep[e.g.][]{Blaizot:2023}, thus explaining the relative shift in our data-points in Figure \ref{fig:fesc_A}.

Figure \ref{fig:fesc_A} shows \fesclos\ as a function of \asym\ for all of our mock observations, coloured by \fx. We also include the simulation data from \cite{Kakiichi:2021} as well as observational data from \cite{Izotov:2016,Izotov:2018b,Izotov:2018} in red and green respectively. We find that our data agrees with the observational data, but that our strong LyC leakers tend to have larger \asym\ than those produced by \cite{Kakiichi:2021}. Furthermore, though we do find evidence for a characteristic transition between the two behaviours, this occurs at a larger red peak asymmetry (at $\sim 6$) than predicted by \cite{Kakiichi:2021} (as indicated by the vertical line on Figure \ref{fig:fesc_A}). Here, it is important to note that not all H~{\small II} regions in \sphinx\ galaxies are fully resolved. In these cases, \asym\ is likely to be reduced as \lya\ emission is surrounded by the uniform medium of an under-resolved cell. 

\subsection{Extended \lya\ Haloes}

\begin{figure}
    \includegraphics{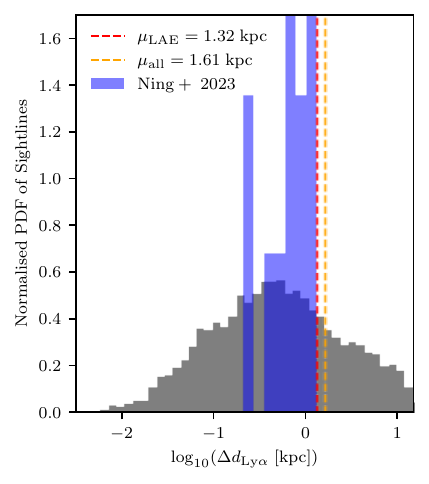}
    \caption{Normalised histogram of the physical offset between the centroids of brightest \lya\ and F150W emission for our mock images. We include observational measurements from \protect\cite{Ning:2023} as a comparison. We find a mean offset of $\SI{1.32}{\kilo\parsec}$ for LAEs and $\SI{1.61}{\kilo\parsec}$ for all galaxies.}
    \label{fig:lya_offsets}
\end{figure}

\begin{figure}
    \includegraphics{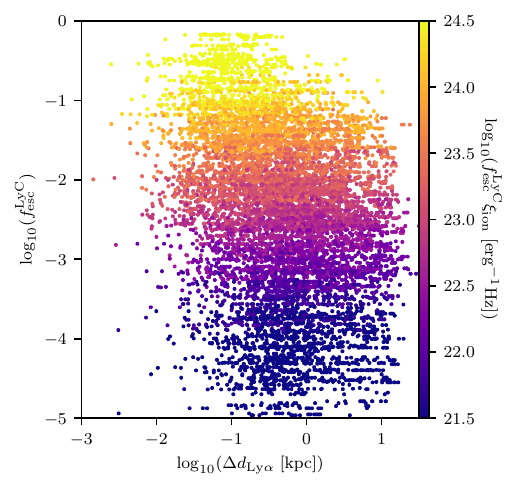}
    \caption{Global LyC escape fraction as a function of the line-of-sight physical offset between the centroids of brightest \lya\ and F150W emission, coloured by \fx. No clear trend is visible, but galaxies which contribute the most to reionization tend to have offsets of $\Delta d_{\rm Ly\alpha} \lesssim \SI{1}{\kilo\parsec}$.}
    \label{fig:fesc_offset}
\end{figure}

\begin{figure*}
    \includegraphics{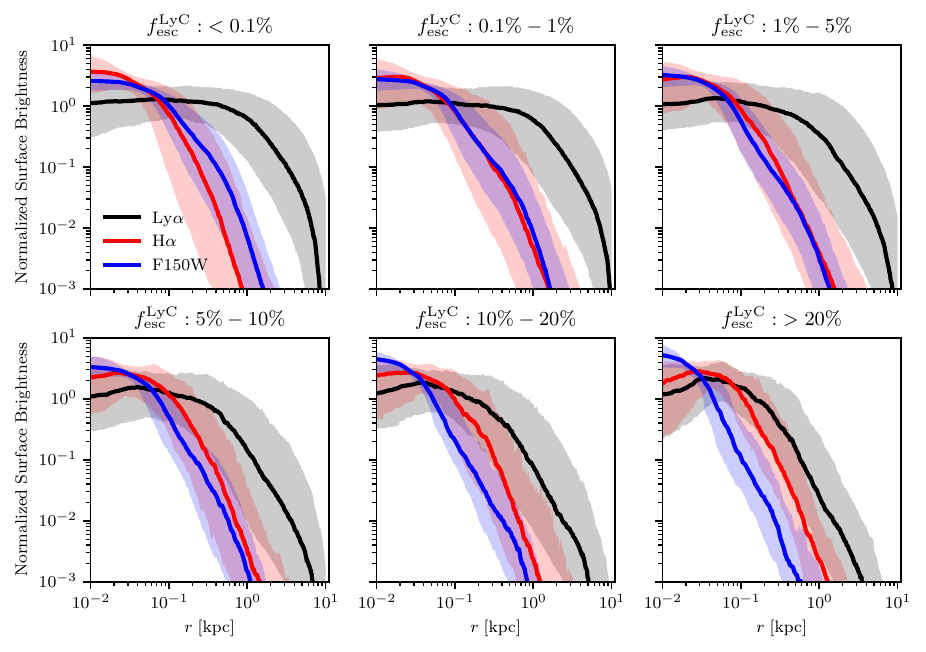}
    \caption{Median normalised surface brightness profiles for \lya\ (black), H$\alpha$ (red), and F150W (blue) images for our mock observations binned by their LyC escape fractions. 16th and 84th percentiles are also shown to demonstrate the spread. We find that in galaxies with significant LyC leakage ($f_{\rm esc}^{\rm LyC} > 20\%$), the \lya\ surface brightness profile is not significantly extended as compared to its H$\alpha$ and F150W profiles, in contrast to stacks of galaxies with low \fesc.}
    \label{fig:lya_prof_fesc}
\end{figure*}

\begin{figure}
    \includegraphics{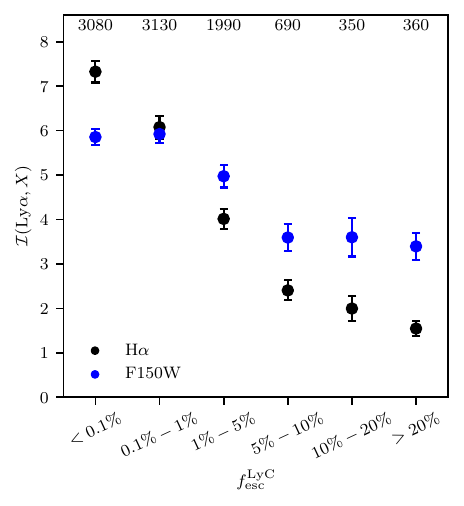}
    \caption{Integral ratios (as defined by Equation \protect\ref{eq:ID}) for median surface brightness profiles in \lya, H$\alpha$, and F150W images of galaxies, binned with respect to their LyC escape fraction. Error bars indicate the standard error of each measurement using the 16th and 84th percentiles as shown in Figure \protect\ref{fig:lya_prof_fesc}. The number of galaxies in each bin is labelled above each point. We find that when stacked, bins of galaxies with higher LyC escape fractions tend to have smaller integral ratios.}
    \label{fig:ID_fesc}
\end{figure}

\begin{figure*}
    \includegraphics{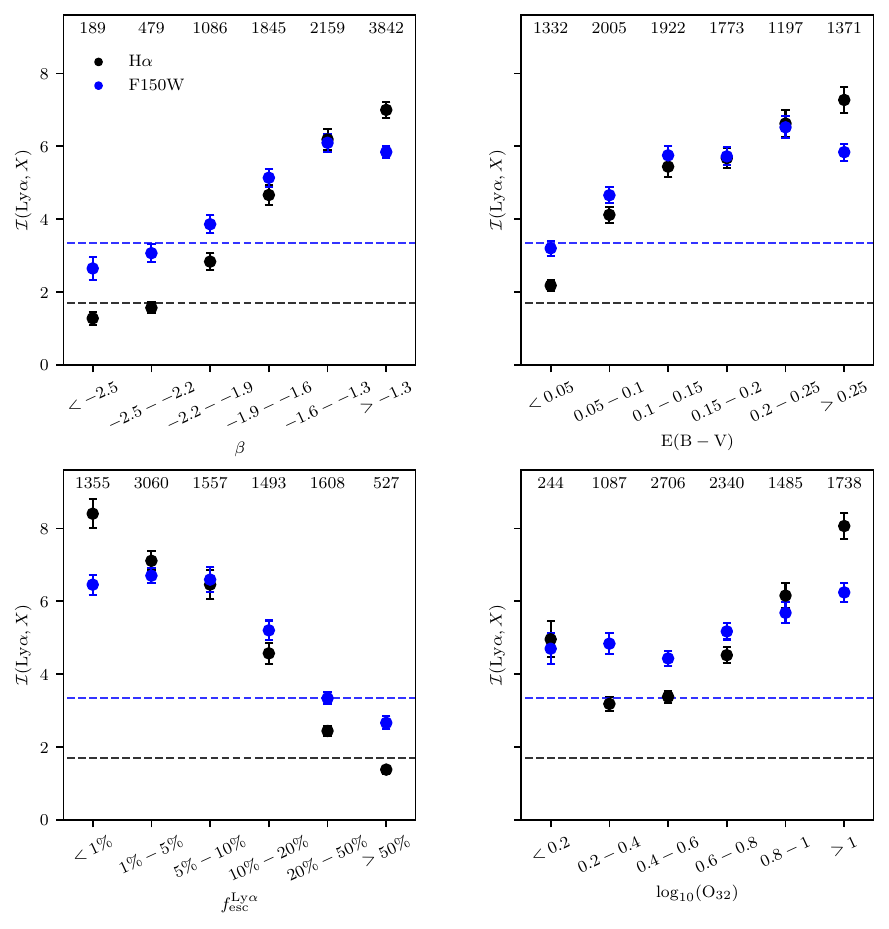}
    \caption{(\textit{Top-left}) As in Figure \ref{fig:ID_fesc}, but for the UV stellar continuum slope, $\beta$. We also indicate the integral ratio values corresponding to bins with $f_{\rm esc}^{\rm LyC} > 10\%$ with horizontal lines. We find a strong correlation between the integral ratio and $\beta$, suggesting that stacks of galaxies with small integral ratios (and thus high LyC escape fractions) tend to have negative UV slopes, in agreement with observations \protect\citep[e.g.][]{Chisholm:2022, Flury:2022b}. (\textit{Top-right}) As above, but for the UV nebular attenuation, $\rm E(B - V)$. We find that bins with smaller integral ratios tend to have less UV attenuation. This agrees with previous work suggesting that galaxies with less dust have higher LyC escape fractions \protect\citep[e.g.][]{Saldana-Lopez:2022}. (\textit{Bottom-left}) As above, but for the \lya\ escape fraction. We find no strong correlation for $f_{\rm esc}^{\rm Ly\alpha} \lesssim 5\%-10\%$, but with a very strong dependency for galaxies leaking more \lya\ photons. As before, this confirms the fact that high \fesca\ is an indicator for LyC leakage. (\textit{Bottom-right}) As before, but for ${\rm O}_{32}$. We find no strong correlation, albeit a weak indication that bins with $0.2 < {\rm O}_{32} < 0.4$ should have the highest LyC escape fractions. This disagrees with the notion that larger ${\rm O}_{32}$ ratios directly correspond to higher \fesc\ \protect\citep[e.g.][]{Nakajima:2014, Izotov:2018b}, though the connection between ${\rm O}_{32}$ and LyC escape is by no means completely understood \citep[c.f.][]{Katz:2020b,Barrow:2020,Flury:2022b}.}
    \label{fig:ID_big}
\end{figure*}

Extended \lya\ haloes (LAHs) have now been measured across a large range of redshifts. For example, studies using the Lyman Alpha Reference Sample \citep[LARS;][]{Ostlin:2014,Hayes:2014} of $z \lesssim 0.2$ galaxies, found a large number of LAHs which extended up to four times further than the effective radii of H$\alpha$, far UV \citep{Hayes:2013,Rasekh:2022} and UV \citep{Yang:2017b} emission. At intermediate redshifts ($z \sim 2.65$), narrow band imaging of Lyman break galaxies (LBGs) has been successful in finding \lya\ profiles exceeding their UV counterparts by a factor of 5-10 \citep{Steidel:2011}. Sensitive integral field spectrographs, e.g. MUSE, have also been leveraged to detect individual LAHs at intermediate redshifts \citep[$2 \lesssim z \lesssim 6$, e.g.][]{Wisotzki:2016,Leclercq:2017,Erb:2018,Leclercq:2020}. JWST has detected extended \lya\ emission at redshifts of $z\sim 7.47-7.75$ \citep{Jung:2023} and $z = 10.6$ \citep{Bunker:2023}. Despite these observations, there is little evidence for evolution in the relative sizes of extended \lya\ or H$\alpha$ (or UV) haloes as a function of redshift \citep[e.g.][]{Runnholm:2023}.

Due to the resonant nature of Ly$\alpha$, the source of extended emission remains unclear. For example, \cite{Kim:2020} used polarimetry to study extended \lya\ emission from a \lya\ nebula at $z = 2.656$ containing an obscured, embedded AGN. They found that escaping \lya\ emission (sourced by AGN-photo-ionized gas) is scattered by the cloud at large radii back into the sight-line. Furthermore, post-processed simulations of star-forming galaxies from the IllustrisTNG50 simulation were used to infer that the majority of photons observed in LAHs are re-scattered photons from star-forming regions \citep{Byrohl:2021}, as opposed to photons emitted in-situ or by satellite galaxies. \cite{Mitchell:2021} also found similar results, studying a cosmological radiation hydrodynamical zoom simulation of a single galaxy. Here, they found that these three emission components contributed equally at a radius of $\sim\SI{10}{\kilo\parsec}$. At very large radii however, these profiles flatten, with the majority of the contributions coming from nearby galaxies, haloes, and cooling radiation rather than diffuse emission. Median-stacked observations at $1.9\lesssim z\lesssim 3.5$ from HETDEX \citep{Niemeyer:2022} as well as at $3\lesssim z\lesssim 4$ from the MUSE Extremely Deep Field \citep{Guo:2023} seem to both confirm this prediction, finding similar surface brightness profiles.

The presence of an extended halo of neutral gas means that \lya\ emission can be scattered, producing larger, more extended LAHs. Moreover, ionizing photons that escape the central source are also able to photoionize neutral hydrogen in the CGM, leading to in-situ emission of \lya\ called fluorescence \citep{Furlanetto:2005,Nagamine:2010}. This process leads to more concentrated, centrally peaked LAHs \citep{Mas-Ribas:2016}. Finally, \lya\ emission can also be powered by gravitational cooling, caused by the accretion of neutral hydrogen onto the galaxy. In contrast, while H$\alpha$ is also emitted predominantly by recombination transitions in starburst-powered H~{\small II} regions and can be produced by fluorescence, it is not a resonant transition. This leads H$\alpha$ haloes to generally be much smaller than their LAH counterparts. The UV continuum is even simpler, as it is only emitted in the ISM surrounding ionizing sources, making it a direct tracer of star formation. All of this, as well as the relative contributions of satellite haloes at larger impact parameters to \lya\ emission \citep{Mas-Ribas:2017} led \cite{Mas-Ribas:2017b} to claim that extended \lya, H$\alpha$, and continuum emission can be used to infer the escape fraction of ionizing radiation from a central source into the CGM. We build on this, using the fact that other \lya-based diagnostics (as discussed above) tend to work by indicating the existence of escape channels for ionizing radiation to suggest that the relative size of extended LAHs is inversely proportional to the global escape of LyC photons.

In order to test this with our simulation, we measure the surface brightness profiles using mock images taken along all 10 lines-of-sight of each galaxy in \lya, H$\alpha$, and UV (using the F150W JWST/NIRCam filter) emission. Because \lya\ emission is not necessarily co-spatial with H$\alpha$ or UV flux, we choose to first re-center all images onto the centroid of the brightest region in the smoothed F150W image, as segmented by {\small PHOTUTILS} \citep{Bradley:2023}. Doing so, we find that a large number of our mock images have a significant offset between the brightest \lya\ and UV emission ($\Delta d_{\rm Ly\alpha}$), in agreement with recent results from \cite{Ning:2023}, who find an average offset of $\Delta d_{\rm Ly\alpha} \sim \SI{1}{\kilo\parsec}$ for a sample of 14 LAEs at $z\sim6$ imaged using JWST/NIRCam. Figure \ref{fig:lya_offsets} shows a normalised histogram of these offsets for all galaxies in our mock sample, as well as those from \cite{Ning:2023}. We find an average offset between \lya\ and F150W emission of $\langle\Delta d_{\rm Ly\alpha}\rangle = \SI{1.32}{\kilo\parsec}$ for galaxies with $W_{\lambda} >25\AA$ and $\SI{1.61}{\kilo\parsec}$ for all galaxies. Figure \ref{fig:fesc_offset} shows the angle-averaged LyC escape fraction as a function of the physical offset between centroids of \lya\ and F150W emission, coloured by \fx. We find that galaxies that contribute most to reionization tend to have $\Delta d_{\rm Ly\alpha} \lesssim \SI{1}{\kilo\parsec}$, but that there is no clear trend between \fesc\ and $\Delta d_{\rm Ly\alpha}$. Following this re-centering process, we then calculate the surface brightness in circular bins, before normalising by the average surface brightness within the central $\SI{10}{\kilo\parsec}$.

We have stacked surface brightness profiles based on the value of \fesc\ for each galaxy. Figure \ref{fig:lya_prof_fesc} shows the median normalized surface brightness profiles of \lya\ (black), H$\alpha$ (red), and F150W (blue), with shaded regions indicating the 16th and 84th percentiles of the respective distributions. \lya\ profiles have no central peak due to the offset discussed above. Galaxies with larger angle-averaged \fesc\ tend to have less extended \lya\ and F150W profiles, while also having more extended H$\alpha$ surface brightness profiles. However, changes in F150W and H$\alpha$ are much smaller than those for Ly$\alpha$. While the physics behind \lya\ profiles has been discussed, it is believed that in LyC leakers, H$\alpha$ profiles become more extended due to fluorescence exciting H$\alpha$ emission in the CGM \citep[e.g.][]{Mas-Ribas:2017}, while the F150W profiles become increasingly steep due to the presence of nuclear starbursts. This indicates the possibility that for galaxies with similar bulk properties, morphological differences might be indicative of LyC leakage.

In order to quantify the relationship between these surface brightness profiles and \fesc, we introduce the integral ratio, $\mathcal{I}({\rm Ly\alpha}, X)$, given by:
\eq{
\mathcal{I}({\rm Ly\alpha}, X) = \frac{\int_{0}^{\SI{5}{\kilo\parsec}} \langle \Sigma_{\rm Ly\alpha}\rangle \;dr}{\int_{0}^{\SI{5}{\kilo\parsec}} \langle \Sigma_{X}\rangle \;dr}, \label{eq:ID}
}
where $\langle\Sigma_{\rm X}\rangle$ is the normalised median surface brightness profile for an image in filter $X$. We note that the \lya\ surface brightness profiles (particularly the \textit{top row} of Figure \ref{fig:lya_prof_fesc}) near $r_{\rm vir}$ should be taken as lower-limits as there are clear edge effects appearing due to the fact that we truncate the radiative transfer at the virial radius. Realistically, to better estimate these profiles out to such distances, it is necessary to simulate \lya\ radiative transfer through a significantly larger volume, which is a computational limitation of our work. As a result, we use an upper limit of $\SI{5}{\kilo\parsec}$ in Equation \ref{eq:ID} to avoid uncertainties due to these effects. We note that while changing this upper limit does affect the value $\mathcal{I}$, it does not strongly affect the trends discussed below. 

Figure \ref{fig:ID_fesc} shows the integral ratio calculated using both H$\alpha$ and F150W with respect to \lya\ for each bin in \fesc. The number of lines-of-sight in each bin is printed above the respective point, along with error bars representing values calculated using the 16th and 84th percentiles. Here, we see clearly that systems with higher \fesc\ have smaller integral ratios, tending to a theoretical value of one (corresponding to the LAH having the same size as the halo imaged in the corresponding wavelength). We find that this trend appears for both H$\alpha$ and F150W, albeit slightly weaker for the latter and with more scatter.

In reality, we are unfortunately unable to use \fesc\ to stack observed galaxies at high redshift. However, given the fact that the stacked integral ratio correlates with \fesc, it can be used as an independent method to investigate the quality of other diagnostics for LyC leakage. Figure \ref{fig:ID_big} shows the average integral ratios for bins in the UV stellar continuum slope, $\beta$ (\textit{top-left}), nebular UV attenuation, $\rm E(B - V)$ (\textit{top-right}), \fesca\ (\textit{bottom-left}), and $[\mathrm{OIII}]\lambda5007/[\mathrm{OII}]\lambda\lambda3726,3728$, ${\rm O}_{32}$ (\textit{bottom-right}) as above in Figure \ref{fig:ID_fesc}. In order to guide the eye, we also include values for the integral ratio corresponding to a bin with $f_{\rm esc}^{\rm LyC} > 10\%$ as horizontal lines. These act as effective thresholds. Here, we can use the fact that lower values of $\mathcal{I}({\rm Ly\alpha}, X)$ correspond to bins with greater \fesc\ to test each diagnostic. In the case of $\beta$, we find that bluer UV slopes smoothly correlate with LyC leakers in agreement with previous works \citep[e.g.][]{Chisholm:2022, Flury:2022b, Choustikov:23}. Next, galaxies with less dust attenuation also have lower integral ratios, corresponding to larger \fesc\ \citep[e.g.][]{Saldana-Lopez:2022, Choustikov:23}. As discussed in Section~\ref{sec:fesca}, galaxies with significant \lya\ leakage tend to also have significant LyC leakage. Next, we can compare integral ratios for each quantity with the threshold lines shown on each figure. In this way, we find that $\beta$ is the best of these diagnostics, as the two stacks with the most negative UV slopes both fell below the threshold for strong LyC leakage. This is in agreement with the results of \cite{Choustikov:23}. Finally, we find that this method suggests that galaxies with  $0.2 \lesssim \log_{10}({\rm O}_{32}) \lesssim 0.6$ tend to have the largest \fesc, with larger ${\rm O}_{32}$ ratios corresponding to larger integral ratios and thus lower LyC escape fractions. We find that these trends also appear when using the F150W images, albeit much weaker.

These findings indicate that there is a wealth of information contained in extended photometric data that (while realistically difficult to obtain at high redshift) can help to understand the efficacy of LyC diagnostics at low and intermediate redshifts.

\subsection{Finding LyC Leakers with \lya}
\label{sec:using_lya}

Having explored how well \lya\ emission properties can be used to select for LyC leakage individually, we now continue to study how these properties can be used in conjunction with other observable quantities. In order to more fairly compare with observations, in this section we follow results from Section \ref{sec:ew} and exclude all simulated galaxies from the analysis that have Ly$\alpha$ EWs $<25$~\angstrom, leaving a reduced sample of 63\% of our \sphinx\ galaxies. We also note that in practice, the ability to measure many of these \lya\ derivative quantities is contingent on the quality of continuum subtraction (particularly in the case of \fcen\, e.g. \citealt{Naidu:2022}). As a result, beginning with such a selection in \wlya\ is a good way to ensure the quality of subsequent inferences.

\begin{figure} 
    \includegraphics{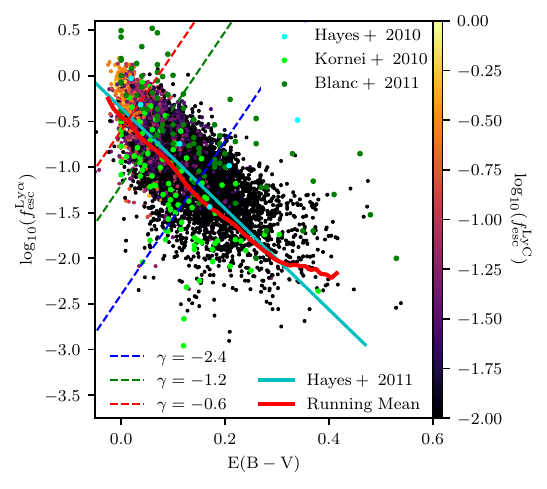}
    \caption{Line-of-sight \protect\lya\ escape fraction as a function of UV attenuation for all mock observations in our sample, coloured by the global LyC escape fraction of the given galaxy. Observational data from \protect\cite{Hayes:2010, Kornei:2010, Blanc:2011} are included in cyan, lime, and green respectively, as well as the best fit from \protect\cite{Hayes:2011} in cyan. Our running mean is shown in red. We include cuts given by Equation \protect\ref{eq:fesca_ebmv_cut} with $\alpha \in {-2.4, -1.2, -0.6}$ as dashed lines in red, green, and blue respectively. The completeness and purity of each cut are given in Table \protect\ref{tab:fesca_ebmv_cut}. We find that strong leakers can be identified by combining these two quantities, with lines-of-sight with large \protect\fesca\ and low $\rm E(B - V)$ being significantly more likely to have high LyC escape fractions.}
    \label{fig:fesca_ebmv_cut}
\end{figure}

So far, we have neglected the contribution of the nebular continuum to the UV slope, $\beta$. In doing so, it is expected that this inclusion tends to redden the UV slope. However, the relative contributions of stellar and nebular continua to $\beta$ can, in principle, be disentangled with SED fitting. Nevertheless, it can be useful to derive a good combined diagnostic for LyC leakage that does not require the stellar and nebular contributions to the continuum to be disentangled. To this end, based on Figure \ref{fig:ID_big}, it is clear that the best combination which suitably covers the three criteria for LyC leakage would be \fesca\ and ${\rm E(B - V)}$. Figure~\ref{fig:fesca_ebmv_cut} shows line-of-sight \lya\ escape fractions as a function of line-of-sight ${\rm E(B - V)}$, coloured by the global LyC escape fraction. We also include observational data from \cite{Hayes:2010}, \cite{Kornei:2010}, and \cite{Blanc:2011} in cyan, lime, and green respectively, as well as the best fit from \cite{Hayes:2011} in cyan. We find very good agreement, as demonstrated by our running mean which is also shown in red. We find that mock observations with exceedingly low levels of UV attenuation (i.e. with ${\rm E(B - V)} \lesssim 0.05$) as well as with the highest \lya\ escape fractions ($f_{\rm esc}^{\rm Ly\alpha} > 30\%$) also have the largest global LyC escape fractions. We have developed a selection criterion given by: 
\eq{
\log_{10}(f_{\rm esc}^{\rm Ly\alpha}) \geq 8.4\ {\rm E(B - V)} + \gamma, \label{eq:fesca_ebmv_cut}
}
where $\gamma$ is a free parameter that can be used to inform the stringency of the cut. More specifically, we find that lower values of $\gamma$ tend to give reduced samples that are more complete with respect to LyC leakers, while also giving lower purities.
\begin{table}
	\centering
	\caption{Percentage completeness and purity of LyC leakers in samples as cut using Equation \ref{eq:fesca_ebmv_cut}. We define completeness as $N(\geq f_{\rm esc}^{\rm LyC}\;|\ {\rm cut})/N(\geq f_{\rm esc}^{\rm LyC})$, while purity is defined as $N(\geq f_{\rm esc}^{\rm LyC}|\ {\rm cut})/N({\rm all}\;|\;{\rm cut})$. We note that galaxies selected by these three cuts are responsible for 76\%, 50\%, and 26\% respectively of all ionizing radiation released into the IGM.}
	\label{tab:fesca_ebmv_cut}
	\begin{tabular}{ccccc} 
		\hline
		 \cellcolor[rgb]{0.5, 0.5, 0.5} & \multicolumn{2}{c}{$f_{\rm esc}^{\rm LyC} \geq 5\%$} & \multicolumn{2}{c}{$f_{\rm esc}^{\rm LyC} \geq 20\%$}  \\
        \hline
        $\gamma$ & \% Completeness & \% Purity & \% Completeness & \% Purity \\
        \hline
        $-2.4$ & 99 & 29 & 100 & 8  \\
        $-1.2$ & 70 & 57 & 92 & 19  \\
        $-0.6$ & 33 & 85 & 68 & 45  \\
		\hline
	\end{tabular}
\end{table}
This is exemplified in Table~\ref{tab:fesca_ebmv_cut} where the percentage completeness and purity are given for $\gamma\in \{-2.4, -1.2, -0.6\}$ with respect to mild LyC leakers ($f_{\rm esc}^{\rm LyC} \geq 5\%$) as well as strong LyC leakers ($f_{\rm esc}^{\rm LyC} \geq 20\%$). These cuts are also shown as dashed lines on Figure \ref{fig:fesca_ebmv_cut}. Here, we find that using $\gamma = -2.4$ gives a sample that is fully complete with respect to both populations, sacrificing purity, being only $29\%$ and $8\%$ complete with respect to the mild and strong LyC leakers respectively. In contrast, using a value of $\gamma = -0.6$ gives a reduced sample that is less complete ($33\%$ and $68\%$ for mild and strong LyC leakers respectively). However, these samples are significantly more pure with respect to LyC leakers, being $85\%$ and $45\%$ pure with mild and strong LyC leakers. We also note that galaxies selected by these three cuts are responsible for 76\%, 50\%, and 26\% respectively of all ionizing radiation released into the IGM. As a result, this suggests that \fesca, $\beta$, and $\rm E(B - V)$ can be used as a powerful combined diagnostic for galaxies with significant LyC leakage, as predicted by the framework proposed in \cite{Choustikov:23} and discussed above. 

\begin{figure}
    \includegraphics{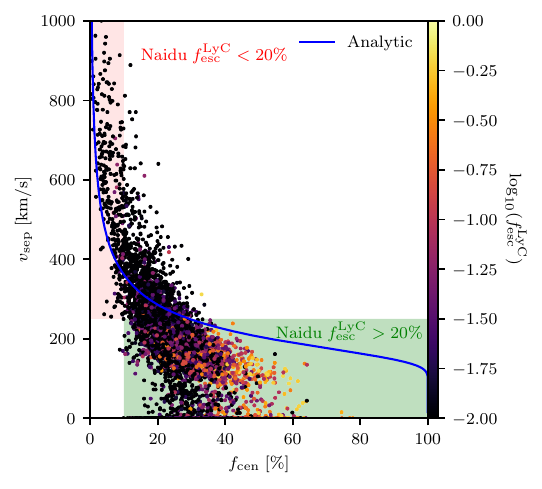}
    \caption{\lya\ peak separation as a function of central escape fractions for all bright LAEs ($W_{\lambda}({\rm Ly\alpha}) > 25$\AA) coloured by the global LyC escape fraction. The simple analytic model from Appendix \protect\ref{sec:homog} is also included. We include suggested selection criteria from \protect\cite{Naidu:2022}, in particular suggesting that $f_{\rm cen} > 10\%$ and $v_{\rm sep} < \SI{250}{\kilo\metre\per\second}$ will select the majority of strong LyC leakers. While this is true (in the case of strong leakers producing a \lya\ doublet), it is still a highly contaminated sample ($94\%$ have $f_{\rm esc}^{\rm LyC} < 20\%$).}
    \label{fig:naidu}
\end{figure}

\cite{Naidu:2022} suggested that galaxies with $f_{\rm cen} > 10\%$ and $v_{\rm sep} < \SI{250}{\kilo\metre\per\second}$ have $f_{\rm esc}^{\rm LyC} > 20\%$. In order to test this, Figure~\ref{fig:naidu} shows \vsep\ as a function of \fcen\ for all of our line-of-sights, coloured by the global LyC escape fraction. Here, the green region indicates the space of galaxies predicted to have large LyC escape fractions, while the red region is reserved for non-leakers. We find that the green region captures $69\%$ of all galaxies with $f_{\rm esc}^{\rm LyC} > 20\%$ in our sample, mainly due to the fact that the remaining strong leakers do not have multiple peaks\footnote{This number increases to $99\%$ if we include lines-of-sight producing single peaks with $f_{\rm cen} > 10\%$.}. However, this region of the observable space is also heavily contaminated by non-leakers, as $94\%$ of these lines-of-sight have $f_{\rm esc}^{\rm LyC} < 20\%$. Thus, we note that while this set of diagnostics certainly works to identify a sample containing the majority of strong LyC leakers (with an average LyC escape fraction of $4.0\%$), we do not reproduce the purity of $80\%$ quoted by \cite{Naidu:2022}. Nevertheless, we include the caveat that there may be a hidden HI mass dependence underlying both of these quantities. To this end, we have also over-plotted the analytic results for the simple analytical model derived in Appendix \ref{sec:homog}, where we discuss this potential selection effect further.

\section{Caveats}
\label{sec:caveats}

The emergent emission spectrum for resonant transmission lines like \lya\ are very sensitive to small scale structures and fluctuations in the density and velocity field of the emitting and attenuating medium. As all cosmological simulations, \sphinx\ has a finite spatial resolution ($\sim10$~pc at $z = 6$). Therefore, while it is able to resolve the multi-phase nature of the ISM, it can not completely capture the small-scale dynamics and feedback processes that are inherent to the ISM or giant molecular clouds \citep[e.g.][]{Kakiichi:2021,Kimm:2022}. While predominantly felt at small scales, these effects (particularly including turbulence, stellar winds, radiation pressure etc.) will change the scattering process of \lya\ and therefore are known to modify the emergent spectral profile in both low and high resolution simulations \citep[e.g.][]{Camps:2021}. Even so, a similar model to that used in this paper has recently been shown to reproduce the plethora of observed galaxy \lya\ spectral profiles \citep{Blaizot:2023}.

Due to the fact that the \sphinx\ simulation does not self-consistently follow the formation and evolution of dust, we use an effective phenomenological dust law, where the dust to metallicity ratio is held constant and dust predominantly traces the neutral gas in our simulation \citep{Laursen:09}. While it is an effective model that has been shown to reproduce observational trends \citep[e.g.][]{Katz:2022, spdrv1, Choustikov:23}, it is important to note the effects that such a model can have, particularly due to the role that dust plays in absorbing \lya\ photons. This differential distribution of dust will have two key affects. The first, is that it may absorb LyC photons before they have had a chance to ionize pockets of neutral hydrogen, thus modifying the spatial distribution of \lya\ emitting gas cells. Secondly, changes to the spatial distribution of Ly$\alpha$-absorbing dust will slightly modify the subsequent \lya\ spectral profiles, particularly in terms of their overall asymmetry (e.g. \citealt{Verhamme:2006}, see also \citealt{Smith2022b}). However, we do not consider this to be a major effect. Finally, the exact dust-to-gas mass relationship is debated, with some observations suggesting that it should follow a power law with metallicity \citep{Remy-Ruyer_2014}. However, studying these effects is beyond the scope of this paper.

Next, while we have included transmission effects due to \lya\ photons travelling through the CGM (i.e. out to the virial radius), we have not included the full effect of the IGM. While the IGM can reduce the overall visibility of LAEs during the Epoch of Reionization \citep[e.g.][see also Appendix \ref{sec:igm}]{Jeeson-Daniel:2012,Behrens:2013,Schenker:2014, Kusakabe2020,Garel:2021}, it can also modify the emergent \lya\ spectral profile, particularly attenuating the blue peak \citep[see Figure 12 of][]{Smith2022}. However, given the fact that we have focused on studying galaxies at redshifts $z \leq 6$, we believe this to only be a minor effect \citep{Garel:2021}. Moreover, our work is also applicable to galaxies contained within large ionized bubbles at higher redshifts \citep[e.g.][]{Saxena:2023b}, as their \lya\ profiles are modified only by transmission through the local ISM and CGM.  

The context and caveats to the physics and emission line modelling behind this \sphinx\ data-set have been explored extensively before. Therefore, we direct interested readers to consult \cite{spdrv1} and \cite{Choustikov:23} for discussions about sub-grid modelling in \sphinx, as well as comparisons to other works utilising simulations such as these.

\section{Summary and Conclusions}
\label{sec:conclusion}

We have post-processed a sample of 960 observable star-forming galaxies at $4.64 \lesssim z \lesssim 6$ from the \sphinx\ cosmological radiation hydrodynamical simulation with {\small CLOUDY} and {\small RASCAS} to produce a library of 9600 resonant-scatter and dust-attenuated \lya\ spectra. We also use {\small RASCAS} to simulate the LyC escape fractions for all 9600 lines-of-sight. We combine these with global LyC escape fractions from the {\small SPHINX} Public Data Release v1 \citep{spdrv1} to carry out the first complete test of the viability of using properties of observed \lya\ emission to infer LyC leakage from epoch of reionization galaxies in a cosmological simulation.

It is confirmed that \lya\ properties of \sphinx\ galaxies are representative of observations of the high-redshift Universe made by JWST. We also found that the typical method of estimating the \lya\ escape fraction produces over-estimates (by as much as two orders of magnitude in extreme cases), particularly for dusty-sight lines where attenuation corrections are sometimes insufficient. 

We have investigated the viability of using spectroscopic properties of \lya\ emission as diagnostics to infer global LyC leakage. The framework for observational diagnostics of LyC leakage proposed by \cite{Choustikov:23} has also been used to explore the physical reasons behind why each diagnostic is successful (or not). This framework states that a good diagnostic for high LyC leakage should select for galaxies with high sSFRs, mean stellar population ages in the range $3.5 \leq \langle\tau_{\rm Stellar}\rangle / [{\rm Myr}] \leq 10$ and should contain a proxy for the density and neutral state of the galaxy's ISM.

Using this we find that the line-of-sight \lya\ escape fraction, \fesca\ is a good diagnostic for LyC leakage, due to the fact that while a weak indicator of sSFR, \fesca\ can be a very good indicator for whether the mean age of a stellar population of a given galaxy is $\gtrapprox \SI{3.5}{\mega\year}$ and unsurprisingly traces the neutral, dusty phase of the ISM well. Next, increased \lya\ equivalent widths, \wlya\ are a weak indicator of sSFR as well as the dust attenuation and neutral gas density of the ISM. However, large \wlya\ do not trace the stellar population age. As a result, by satisfying 2/3 criteria weakly, we find that \wlya\ is a necessary but insufficient diagnostic for LyC leakage. Next, large \lya\ peak separations, \vsep\ were found to select for stellar populations too young to clear channels in their ISM, correlating with UV attenuation and neutral gas density. As a result, we find that strong LyC leakers tend to have $v_{\rm sep} < \SI{250}{\kilo\metre\per\second}$. However, given the fact that \vsep\ does not correlate with sSFR, we find that this is a necessary but insufficient diagnostic for LyC leakage. The fraction of \lya\ photons escaping near line centre, \fcen\ was found to correlate strongly with the density of the dusty ISM. Furthermore, we find that for $f_{\rm cen} \gtrsim 30\%$, \fcen\ correlates with sSFR and for $f_{\rm cen} \gtrsim 40\%$ selects galaxies with mean stellar population ages in the correct range for effective LyC leakage. As a result, we find that \fcen\ has the possibility of being a very useful diagnostic for LyC leakage, albeit with the caveat that the $f_{\rm cen} - f_{\rm esc}^{\rm LyC}$ relationship is far from trivial as there is likely a hidden galaxy mass dependence. Finally, while the asymmetry of the red peak, \asym\ has been explored as a useful tool for investigations of the exact method of LyC leakage on small scales, we find that it does not correlate with sSFR or the mean stellar population age. Interestingly, we find that the strongest leakers tend to be clustered around $A_f^{\rm red} \sim 3$ (with a slight bias towards larger values) due to the fact that such lines-of-sight tended to have less dense or dusty ISMs. However, given the fact that \asym\ only marginally informs us of 1/3 criteria, we conclude that it is an unsuitable indicator for LyC leakage by itself. 

Building on the work of \cite{Mas-Ribas:2017b}, we have used cosmological simulations to investigate the connection between extended \lya, H$\alpha$, UV continuum (F150W) emission, and LyC escape. Studying re-centred and stacked mock images of our galaxies at these wavelengths, we find that strong LyC leakers tend to have contracted \lya\ and UV haloes with similar surface brightness profiles to their H$\alpha$ haloes (which in contrast are slightly more extended). In contrast, stacked samples with significantly extended \lya\ haloes tend to have low LyC escape fractions. This follows from the fact that the majority of extended \lya\ emission is believed to be re-scattered light from the central star-forming regions as well as fluorescence, implying the significant presence of neutral hydrogen in the CGM. Using the integral ratio as defined in Equation \ref{eq:ID}, we have also explored how stacked \lya\ profiles compare to their H$\alpha$ and F150W counterparts when stacked in bins according to other potential \fesc\ diagnostics, including the UV slope, $\beta$, UV attenuation, $\rm E(B - V)$ and O$_{32}$. We find that this method independently verifies previous results that $\beta$ and $\rm E(B - V)$ are good diagnostics for \fesc, while O$_{32}$ is a necessary but insufficient diagnostic \citep{Choustikov:23}. This exercise confirms the fact that \lya\ surface brightness morphology can be used to understand the leakage of ionizing radiation from the centres of galaxies.

Finally, the possibility of using properties of \lya\ emission to infer large LyC escape fractions was also explored. Given the discussions above, it was found that \fesca\ is the most promising feature, despite the fact that it is often over-predicted in observational studies (see Section \ref{sec:lya_props}). A combined criterion which should be unaffected by pollution from nebular continuum emission is proposed by combining \fesca\ with $\rm E(B - V)$. Here, the cut given in Equation~\ref{eq:fesca_ebmv_cut} is found to provide a flexible method to select LyC leaker-enriched samples with desired completeness and purity. In general, it is found that a combination of \fesca, $\beta$ and $\rm E(B - V)$ are best at selecting for LyC leaking galaxies.

We have explored the feasibility of various \lya-based indirect diagnostics for galaxies with high LyC escape fractions. By using a rich data-set of mock \lya\ observations of simulated high-redshift galaxies with multi-phase ISMs, as well as a physically-motivated theoretical framework for the physics driving LyC leakage, we have found that properties of \lya\ spectral and surface brightness profiles can indeed be used as reliable tracers for LyC leakage. This is in agreement with recent observational studies of the local and high-redshift Universe. We recognise that these results depend on the resolution limits of \sphinx\ as well as the sub-grid physics used, despite the fact that it is a state-of-the-art simulation of galaxy formation during the Epoch of Reionization. Nevertheless, this work highlights the potential of JWST data to find and understand the sources of cosmological reionization, further deepening our understanding of the cosmic dawn.

\section*{Acknowledgements}

N.C. thanks Alex J. Cameron and Sophia Flury for insightful discussions. N.C. and H.K. also thank Jonathan Patterson for helpful support on Glamdring throughout the project.

N.C. acknowledges support from the Science and Technology Facilities Council (STFC) for a Ph.D. studentship. HK acknowledges support from the Beecroft Fellowship. T.K. is supported by the National Research Foundation of Korea(NRF) grant funded by the Korea government(MSIT) (2020R1C1C1007079 and 2022R1A6A1A03053472).

This work used the DiRAC@Durham facility managed by the Institute for Computational Cosmology on behalf of the STFC DiRAC HPC Facility (www.dirac.ac.uk). The equipment was funded by BEIS capital funding via STFC capital grants ST/P002293/1, ST/R002371/1 and ST/S002502/1, Durham University and STFC operations grant ST/R000832/1. DiRAC is part of the National e-Infrastructure.This work was performed using the DiRAC Data Intensive service at Leicester, operated by the University of Leicester IT Services, which forms part of the STFC DiRAC HPC Facility (www.dirac.ac.uk). The equipment was funded by BEIS capital funding via STFC capital grants ST/K000373/1 and ST/R002363/1 and STFC DiRAC Operations grant ST/R001014/1. DiRAC is part of the National e-Infrastructure.

Computing time for the SPHINX project was provided by the Partnership for Advanced Computing in Europe (PRACE) as part of the ``First luminous objects and reionization with SPHINX (cont.)''  (2016153539, 2018184362, 2019215124) project. We thank Philipp Otte and Filipe Guimaraes for helpful support throughout the project and for the extra storage they provided us. We also thank GENCI for providing additional computing resources under GENCI grant A0070410560.  Resources for preparations, tests, and storage were also provided by the Common Computing Facility (CCF) of the LABEX Lyon Institute of Origins (ANR-10-LABX-0066) and PSMN (Pôle Scientifique de Modélisation Numérique) at ENS de Lyon.

\section*{Author Contributions}

The main roles of the authors were, using the CRediT (Contribution Roles Taxonomy) system\footnote{\url{https://authorservices.wiley.com/author-resources/Journal-Authors/open-access/credit.html}}:

\textbf{Nicholas Choustikov}: Conceptualization; Formal analysis; Writing - original draft; Methodology; Visualisation. \textbf{Harley Katz}: Conceptualization; Formal analysis; Writing - original draft; Methodology. \textbf{Aayush Saxena}: Conceptualization; Writing - review and editing. \textbf{Thibault Garel}: Software; Writing - review and editing. \textbf{Julien Devriendt}: Resources; Supervision; Writing - review and editing. \textbf{Adrianne Slyz}: Resources; Supervision; Writing - review and editing. \textbf{Taysun Kimm}: Writing - review and editing. \textbf{Jeremy Blaizot}: Writing - review and editing. \textbf{Joki Rosdahl}: Writing - review and editing.

%%%%%%%%%%%%%%%%%%%%%%%%%%%%%%%%%%%%%%%%%%%%%%%%%%
\section*{Data Availability}
The SPHINX$^{20}$ data used in this article is available as part of the SPHINX Public Data Release v1 \citep[SPDRv1,][]{spdrv1}.

%%%%%%%%%%%%%%%%%%%% REFERENCES %%%%%%%%%%%%%%%%%%

\bibliographystyle{mnras}
\bibliography{References.bib}

%%%%%%%%%%%%%%%%%%%%%%%%%%%%%%%%%%%%%%%%%%%%%%%%%%

%%%%%%%%%%%%%%%%% APPENDICES %%%%%%%%%%%%%%%%%%%%%
\appendix

\section{IGM Attenuation}
\label{sec:igm}
\begin{figure*}
\includegraphics{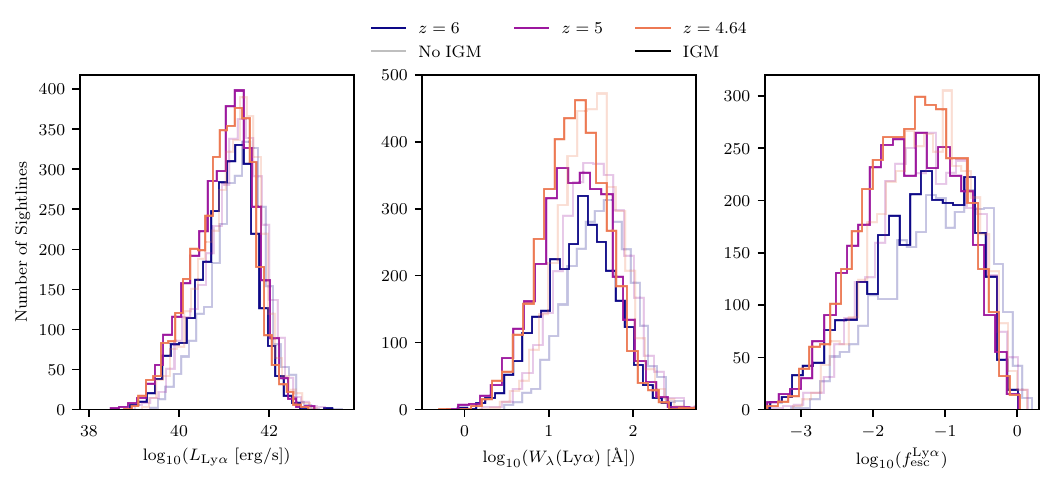}
    \caption{Histograms of \lya\ luminosity (\textit{left}), equivalent width (\textit{center}) and escape fraction (\textit{right}) with (\textit{solid}) and without (\textit{faint}) IGM attenuation, using the analytic expression for IGM optical depths from \protect\cite{Inoue:2014}.}
    \label{fig:lya_igm}
\end{figure*}

In Section \ref{sec:lya_props}, we explored the \lya\  properties of our sample of mock observations, including comparisons to JWST observations. Due to the fact that these galaxies are observed during the epoch of reionization, it is important to account for absorption due to neutral hydrogen in the IGM.

To explore the effect of including IGM attenuation, we apply the analytic model for the transmission function of the IGM given in \cite{Inoue:2014} to our \lya\ spectra and re-compute the, \lya\ luminosities, equivalent widths, and escape fractions. In Figure \ref{fig:lya_igm} we show histograms of each of these quantities, comparing values with (\textit{solid}) and without (\textit{faint}) the IGM correction. We find that as expected, the effect of IGM attenuation is to reduce each of these quantities. However in practice, these IGM-attenuated observables are still entirely consistent with observations shown in Figure \ref{fig:lya_pop_histograms} \citep{Saxena:2023,Roy:2023}.

\section{line-of-sight vs. Global LyC escape fractions}
\label{sec:aa_los}

It is clear that the processes leading to leakage of LyC radiation are profoundly chaotic and depend on local processes of galaxy evolution. As a result, the amount of ionizing radiation which escapes has a strong line-of-sight dependence \citep[e.g.][]{Gnedin:2008,Zackrisson:2013,Fletcher:2019, Kimm:2022, Katz:2022}. As a result, it is important to understand whether line-of-sight measurements of LyC leakage are representative of the global escape fraction for a given galaxy. This is particularly crucial given the fact that observational studies which are able to directly measure LyC emission from galaxies \citep[e.g.][]{Flury:2022a} are limited by only being able to observe galaxies from a single perspective. As a result, their samples may be polluted by fortuitous lines-of-sight with uncharacteristically high, or low, LyC escape fractions.  

\begin{figure}
\includegraphics{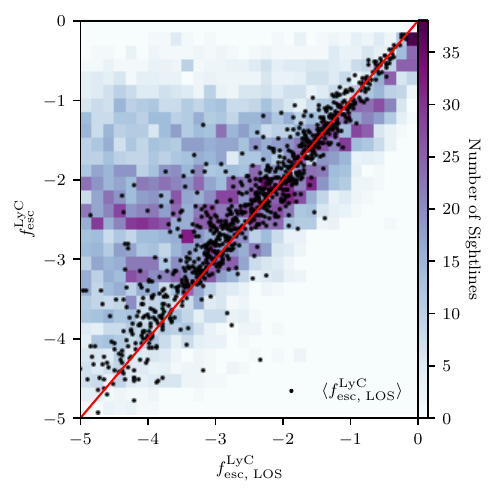}
    \caption{Histogram of global LyC escape fractions, \fesc\ against the 10 line-of-sight LyC escape fractions, \fesclos\ measured for each galaxy. We also include the angle averaged line-of-sight values for each true global value in black. The one-to-one relation is shown in red.}
    \label{fig:fesc_aa_los}
\end{figure}

To this end, we compare the global LyC escape fractions, \fesc\ to the 10 line-of-sight LyC escape fractions, \fesclos\ measured for each galaxy in our mock sample. Figure \ref{fig:fesc_aa_los} shows a histogram of \fesc\ versus all 10 \fesclos\ values for each galaxy. We also include the angle averaged line-of-sight values for each galaxy in black, finding that while we are still affected by small number statistics for a single galaxy, we do recover the expected one-to-one relation (shown in red) for the full mock sample. Interestingly, we find that galaxies with the largest global escape fractions (i.e. $\gtrsim 5\%$) tend to show more isotropic leakage. On the other hand, galaxies with intermediate global leakage (i.e. $\sim 1\%$) tend to be dominated by a few channels with effective leakage.

\section{Offsets due to H~{\small I} Mass of \sphinx\ Galaxies}
\label{sec:homog}

In Section \ref{sec:results}, we found that compared to observational data from the LzLCS \citep{Flury:2022b} and \cite{Naidu:2022}, the \lya\ spectra of \sphinx\ galaxies tended to exhibit systematically lower \vsep\ and higher $f_{\rm cen}$. Here we explore the origin of this discrepancy. Similar discrepancies were found in cosmological zoom-in simulations of a dwarf galaxies by \cite{Yuan:2024}.

It has already been established that \sphinx\ galaxies are less massive and less UV-bright than those discussed by both the LzLCS and \cite{Naidu:2022}. As a result, it is reasonable to assume that they will have smaller H~{\small I} masses \citep{Parkash:2018} and therefore lower H~{\small I} column densities. 

In order to understand the impact of H~{\small I} column density on \fcen\ and $v_{\rm sep}$, it is instructive to use a simple analytical model. We consider a source of \lya\ at the centre of a homogeneous, neutral spherical cloud, with optical depth $\tau_0$. The emergent normalised spectrum has the form \citep{Dijkstra:2006}
\eq{
J(x) = \frac{\pi^{3/2}}{\sqrt{6}a\tau_0} \ls \frac{x^2}{1 + \cosh{\lb\sqrt{\frac{2\pi^3}{27}} \frac{|x^3|}{a\tau_0}\rb}}\rs,
}
where $x = (\nu - \nu_0)/\Delta \nu_D$ is the dimensionless frequency relative to the \lya\ line centre at $\num{2.47e15}\;{\rm Hz}$, $\Delta \nu_D = v_{\rm th} \nu_0 / c$ for the thermal velocity $v_{\rm th} = \sqrt{2k_B T/m_p}$. Here, $k_B$ is the Boltzmann constant, $T$ is the gas temperature, and $m_p$ is the proton mass. Finally, $a = A_{21}/4\pi \Delta v_D$ is the Voigt parameter, where $A_{21}$ is the Einstein $A$-coefficient for the transition. Furthermore, under these assumptions
\eq{
\tau_0 = N_{\rm HI} \sigma_{0} \approx \num{8.3e6} \lb \frac{N_{\rm HI}}{\num{2e20}\;{\rm cm}^{-2}}\rb \lb \frac{T}{\num{2e4}\;{\rm K}}\rb,\label{eq:tau}
}
allowing us to connect this theoretical spectrum to the H~{\small I} column density discussed before. Based on this analytical expression (which we note is symmetric about $x = 0$), we are able to derive expressions for \vsep\ and \fcen\ directly as a function of $N_{\rm HI}$. Doing so, we find
\eq{
v_{\rm sep} = \frac{2c u}{1 - u^2},\;\; u = \frac{x_p \Delta\nu_D}{\nu_0},\;\;x_p = 0.92(a \tau_0)^{1/3},\label{eq:vsep}
}
and
\eq{
f_{\rm cen} = \tanh{\ls \frac{\pi^{3/2}}{3\sqrt{6}a\tau_0}\lb\frac{\nu_0}{\Delta\nu_D}\frac{V_{\rm cut}}{c - V_{\rm cut}}\rb^3\rs},\label{eq:fcen}
}
where we use a cut-off at $V_{\rm cut} = \SI{100}{\kilo\metre\per\second}$ to compare to the method used by \cite{Naidu:2020}. We note that these expressions only hold for the static, simple geometry that we have considered here.

\begin{figure}
\includegraphics{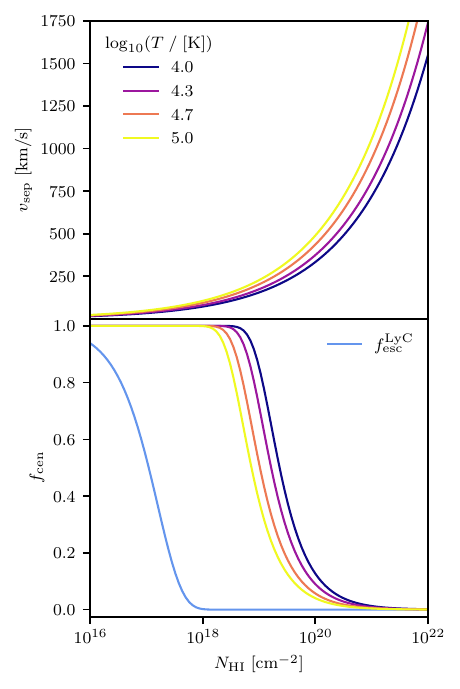}
    \caption{\protect\lya\ peak separation, \protect\vsep\ and central escape fraction, \protect\fcen\ for a central source in a static, homogeneous sphere of gas, as given by Equations \protect\ref{eq:vsep} and \protect\ref{eq:fcen}.}
    \label{fig:vsep_fcen_nhi}
\end{figure}

Figure \ref{fig:vsep_fcen_nhi} shows \vsep\ and \fcen\ as functions of $N_{\rm HI}$ calculated using Equations \ref{eq:vsep} and \ref{eq:fcen} for various gas temperatures. We also include the LyC escape fraction for this simplified setup to help guide the eye. It is clear that both \vsep\ and \fcen\ can vary strongly with H~{\small I} column density. For the \sphinx\ data-set, we have average values of $\left<v_{\rm sep}\right> = 252\ {\rm km}/{\rm s}$ and $\left<f_{\rm cen}\right> = 26\%$ respectively. Inverting Equations \ref{eq:vsep} and \ref{eq:fcen}, these correspond to an H~{\small I} column density of $\sim10^{19.5} - 10^{20}\ {\rm cm}^{-2}$. Furthermore, an average H~{\small I} column density only 2.5 times greater than that of the \sphinx\ average would correspond to a \vsep\ of $\sim350\ {\rm km}/{\rm s}$ and \fcen\ of $\sim 10\%$, consistent with the data from the LzLCS \citep{Flury:2022b} and \cite{Naidu:2022}. Indeed, in Figure \ref{fig:naidu} we can see that even such a simple model is able to capture the crux of the relationship between two such complex quantities reasonably well. Therefore, the discrepancies discussed above are likely caused by differences in the H~{\small I} column densities of these observations compared to the \sphinx\ simulations. 

As a side-note, it is also clear from Figure \ref{fig:vsep_fcen_nhi} that while this simple model can be instructive, it is clearly insufficient to capture the intricacies of the systems being studied. Namely, this model predicts that \vsep\ will only be a viable diagnostic below H~{\small I} column densities of $\sim10^{18}\ {\rm cm}^{-2}$ and that \fcen\ could never be a diagnostic for LyC escape, in contrast to the findings of Section \ref{sec:results}. Therefore, we can conclude that these two diagnostics function best when the ISM is in-homogeneous and contains dust.

% Don't change these lines
\bsp	% typesetting comment
\label{lastpage}
\end{document}